\documentclass[aps,preprint,12pt]{article}
\usepackage{graphicx}
\usepackage{amssymb}
\usepackage{epstopdf}
\usepackage[T1]{fontenc}

\usepackage{setspace}
 \title{The power of a good idea: 
 quantitative modeling of the spread of ideas from epidemiological models}
 \author{  Lu\'{i}s M. A. Bettencourt$^1$, Ariel Cintr\'{o}n-Arias$^{2,3}$ 
\footnote{corresponding author: ariel@cam.cornell.edu } 
 , \\ 
David I. Kaiser$^4$  and Carlos Castillo-Ch\'{a}vez$^{2,3}$   \\ 
 \\ $^1$CCS-3, Computer and Computational Sciences, 
 \\ Los Alamos National Laboratory, Los Alamos NM 87545 \\ 
$^2$Center for Applied Mathematics, Cornell University \\ 657 Rhodes 
Hall, Ithaca NY 14853  \\
$^3$Department of Mathematics and Statistics, Arizona State University  \\ 
PO BOX  871804.  Tempe, AZ - 85287 - 1804 \\
$^4$Center for Theoretical Physics and Department of Physics, \\
Massachusetts Institute of Technology,  \\
77 Massachusetts Ave., Cambridge MA 02139 }

\begin{document}
\maketitle
\date

{\hfill Preprint Number: LAUR-05-0485, MIT-CTP-3589}
\vskip 0.5 cm	

\newpage

\abstract{The population dynamics underlying the diffusion of ideas hold
many qualitative similarities to those involved in the spread of
infections.  In spite of much suggestive evidence this analogy is hardly
ever quantified in useful ways. The standard benefit of modeling epidemics
is the ability to estimate quantitatively population average parameters,
such as interpersonal contact rates, incubation times, duration of
infectious periods, etc. In most cases such quantities generalize
naturally to the spread of ideas and provide a simple means of quantifying
sociological and behavioral patterns. Here we apply several paradigmatic
models of epidemics to empirical data on the advent and spread of Feynman
diagrams through the theoretical physics communities of the USA, Japan, and 
the USSR in the period immediately after World War II. This test case has
the advantage of having been studied historically in great detail, which
allows validation of our results.  We estimate the effectiveness of
adoption of the idea in the three communities and find values for parameters
reflecting both intentional social organization and long lifetimes for the
idea. These features are probably general characteristics of the spread of
ideas, but not of common epidemics. }

\newpage

\section{Introduction}
\label{intro}

Dynamical population models are used to predict average behavior, 
generate hypotheses or explore mechanisms across many fields of science including ecology \cite{may, yod, thi},  epidemiology \cite{anma, dihe, alli, brca}  and 
immunology \cite{penemaleho}, to name but a few.  Traditionally, epidemiological models
focus on the dynamics of ``traits'' transmitted between individuals,
communities, or regions (within specific temporal or spatial scales).  Traits 
may include (i) a communicable disease
such as measles \cite{may} or HIV \cite{hucoca}; (ii) a cultural characteristic such
us a religious belief, a fad \cite{watt,bihiwe,bet,stma}, an innovation
\cite{rog}, or fanatic behavior \cite{caso}; (iii) an addiction such us
drug use \cite{sancas} or a disorder \cite{ghovk}; or (iv) information
spread through, e.g., rumors \cite{rap,dake1}, email messages \cite{adhu},
weblogs \cite{adad}, or peer-to-peer computer networks \cite{kekle}.

The earliest and by now most thoroughly studied population models are
those used to map disease progression through a human population
\cite{kemc,het,mopave}.  These models typically divide a
population into classes that reflect the epidemiological status of
individuals (e.g. susceptible, exposed, infected, etc),  who in turn transit 
between classes via mutual contact at given average rates.  
In this way the models can capture average disease progression
by tracking the mean number of people who are infected, 
who are prone to catch the disease, and who have recovered over time.  In addition, these
models can be used to identify the role of specific population characteristics 
such as  age, variable infectivity, and variable infectious periods  \cite{het}. The division of epidemiological classes according to such characteristics gives rise to more complex models with so called heterogeneous mixing.

In this paper we apply models similar to those used in epidemiology to the spread of ideas. By the term ``idea'' we refer generally to any concept that can be transmitted from
person to person \cite{ros0, bod1,brre, keklta}. It may refer to a
technology, which may require effort and apprenticeship to be learned, but
it may also be a more fickle piece of information such as a colloquialism
or a piece of news. What is important is that it is possible to tell if
someone has adopted the idea, understands and remembers it, and is capable
of and/or active in spreading it to others.

Pioneering contributions to the modeling of social contagion
processes, based on epidemiological models, date back to 1953 \cite{rap}.  
Nearly a decade later, models were applied to the spread
of scientific ideas \cite{gof, tab}.  Around the same time, a 
stochastic model for the spread of rumors was proposed 
and analyzed \cite{dake1}.  In this later model, a closed population is
divided into three ``social'' states:  ignorant, spreaders, and stiflers.  
Transitions from the ignorant state to spreaders may result from contacts
between the two classes, whereas encounters between individuals who
already know the rumor may lead to its cessation.  Various recent extensions
of this model include a general class of Markov processes for generating
time-dependent evolution \cite{dipe}, and studies of the effects of social
landscapes on the spread, either through Monte Carlo simulations over
small-world \cite{zan1} and scale-free \cite{mnp} networks, or by
derivation of mean-field equations for a population with heterogeneous
ignorant and spreader classes \cite{tcd}. Despite this revival in the
modeling of information spread, few of these models have been directly
applied to empirical data. In our opinion, this constitutes a serious gap
in the literature, because only the analysis of real data can ultimately
validate model assumptions or point to novel features of such a complex
process.  The main objective of this paper is to bridge this gap.

Beyond obvious qualitative parallels there are also important differences
between the spread of ideas and diseases. The spread of an idea, unlike a
disease, is usually a conscious act on the part of the transmitter and/or
the adopter. Some ideas that take time to mature, such as those requiring
apprenticeship or study, require active effort to acquire.  There is also
no simple automatic mechanism -- such as an immune system -- by means of
which an idea may be cleared from an infected individual. Most
importantly, it is usually advantageous to acquire new ideas, whereas this
is manifestly not so for diseases. This leads people to adopt different,
often opposite, behaviors when interested in learning an idea compared to
what they may do during an epidemic outbreak. Thus we should expect
important qualitative and quantitative differences between ideas and
diseases when using epidemiological models in a sociological context.  We
explore some of these points below in greater detail, in the context of
specific models and data.

In spite of these differences, quantifying how ideas spread is very
desirable as a means of testing sociological hypotheses.  For example, we
can apply dynamical population models to the spread of an idea to validate
statements about how effectively it is transmitted, the size of the
susceptible population, the speed of its spread, as well as its
persistence. Estimating the population numbers and rates is useful in
constraining explanatory frameworks.  It is also useful for studying how
cultural environments may affect adoption, as happens when the same idea
is presented to communities in different nations, or conversely when
different ideas are presented to the same community.

We pursue these goals in this paper by applying several generic models of
epidemiology to the diffusion of a specific scientific idea in three
different communities.  Our test case is the spread of Feynman diagrams,
since the late 1940s the principal computational tool of theoretical
high-energy physics, and later also used extensively in other areas of
many-body theory such as atomic physics and condensed-matter theory.  The
primary reason to choose this example is that we have detailed historical
information about the network of contacts, person by person, by means of
which the diagrams spread during the first six years after their
introduction \cite{DKDTA,KIH,BettencourtKaiser}.

This example of the spread of an idea may not transcend automatically to
other cases of idea diffusion. Feynman diagrams are primarily a tool for
complex calculation. As such their study and assimilation require a period of apprenticeship and
familiarization. Transmission of the technique almost invariably
proceeded, in the early years, through personal contact, from informal
teacher to student and among peer groups of users. In later years the idea
became familiar and available in accessible forms so that (in
principle) it could more easily have been learned from books and lecture
notes.  Thus, although our example will clearly not cover every class of
ideas it will point, we believe, to features of epidemic models that apply
to idea diffusion. It will also reveal features of these models that
require modification, thereby producing more realistic candidate models
that we expect will prove useful beyond our present analysis.

In Section \ref{data description} we give some historical background on
the spread of Feynman diagrams in the United States, Japan, and the Soviet
Union. We discuss our data sources and the organization of the datasets.
Section \ref{models} presents several classes of models of epidemiology
(or directly inspired by them), some of their mathematical properties, and
the circumstances under which we expect them to apply to the spread of
ideas. We apply each model to the historical data in Section
\ref{results}, and discuss the estimated values for the model parameters
in the light of our independent knowledge of how the diagrams spread.
Finally in Section \ref{conclusions} we present our conclusions and give
some outlook on the general population modeling of the spread of ideas.  
Appendix A contains details about our parameter estimation procedure.

\section{Data sources, time series reconstruction, and state determination}
\label{data description}

Feynman diagrams occupy a central role in modern theoretical physics.
Realistic models of high-energy physics, as well as in condensed-matter,
atomic, and nuclear physics cannot be solved exactly to generate
predictions that can be confronted with experiments. In special
circumstances, however, such as when interactions are weak, series
expansions in a small parameter permit very good systematic
approximations.

In models of particle physics, such as the relativistic quantum theory of
electromagnetism -- quantum electrodynamics -- most terms of this series
beyond zeroth order (tree level) are formally infinite.  The procedure of
removing unphysical infinities to generate predictions is called
``renormalization.'' It is vital for renormalization to work that
commensurate terms be grouped together. This is a relatively simple
procedure for the lowest orders in the expansion series but becomes
absolutely confounding at higher orders, in which many terms contribute
and infinities must cancel precisely between them.  For example, in
quantum electrodynamics, second-order calculations (involving the first
non-trivial corrections within the perturbative expansion) typically
involve ten or so distinct terms to be delimited, calculated, and added
together, while eighth-order calculations involve nearly one thousand such
terms.  Both challenges to making calculations in quantum electrodynamics
-- the presence of infinities and the accounting difficulties of
perturbative calculations -- were well known to physicists during the
1930s, and the problems remained unsolved after World War II.  Throughout
1947 and 1948 several approaches to rendering quantum electrodynamics
well-defined were being attempted in the USA and Japan, but it remained
unclear if any renormalization program could succeed systematically
\cite{Schweber}.

It was then that Freeman Dyson, following up on an idea by Richard
Feynman, was able to show how a diagrammatic representation of particle
interactions could be used to organize the series expansion.  Using the
diagrams, Dyson further demonstrated that the infinities could be
systematically identified and cancelled to any perturbative order.  This
conceptual breakthrough unified Feynman's approach (then at Cornell
University) with that of Julian Schwinger (at Harvard University) and
Sin-Itiro Tomonaga (at Tokyo Education University).  For their
contributions Feynman, Schwinger, and Tomonaga were awarded the Nobel
Prize in 1965 \cite{Schweber}. Feynman diagrams opened the floodgates for
computation (and prediction) in quantum electrodynamics and beyond,
creating enormous research opportunities for a new generation of
theoretical physicists. Tests of quantum electrodynamics and later quantum
field theories of the weak and strong nuclear interactions continue today
in multibillion-dollar particle accelerators at CERN and Fermilab, as well
as at smaller installations. These quantum field theories taken together
constitute the ``standard model of particle physics,'' which summarizes
our most fundamental (and most exact) understanding of matter and
radiation to date.  Almost all quantitative predictions of the standard
model, on which modern particle physics and cosmology are based, are
computed using series of Feynman diagrams.

Because of their extraordinary importance in enabling a good part of
modern theoretical physics, the advent of quantum electrodynamics and of
Feynman diagrams in particular has been very well documented. Our data was
collected in large part for a new book by one of the authors \cite{DKDTA}.
For the United States and Britain one of us (Kaiser) reconstructed the
network of contacts -- author by author -- for the spread of the diagrams
during the first six years after their introduction, between 1949 and
1954. For this he relied upon unpublished correspondence, preprints,
lecture notes, and publications from the period, along with more recent
interviews and published recollections.  With the aid of two colleagues,
he used similar materials to study how the diagrams spread to young
physicists in Japan and the Soviet Union.  Although less information is
readily available about these communities of physicists, a reasonably
complete picture of contacts and spread can also be inferred
\cite{DKDTA,KIH}.

Data for the number of authors adopting Feynman diagrams were collected for the first six years in the USA and Japan, from the beginning of 1949 to the end of 1954. For the Soviet Union, where the diagrams were introduced later and where the spread was initially slower, we assembled data for the first eight years, from the beginning of 1952 to the end of 1959.  We identified adopters of the idea (or members of the ``infected'' class) based on published uses of (or discussion of) Feynman diagrams in the main physics research journals of each country: {\it Physical Review} in the USA, {\it Progress in Theoretical Physics} in
Japan, and {\it Zhurnal Eksperimental'noi i Teoreticheskoi Fiziki} ({\it
Journal of Experimental and Theoretical Physics}) in the USSR. 
The data were identified by manual page-by-page counts.  We found this to be necessary because no citation search or even keyword search would suffice.  Often in the early years authors would cite the Feynman and/or Dyson papers without making any use of the actual diagrammatic techniques, and, conversely, by the early 1950s many would use the diagrams without necessarily citing the Feynman or Dyson papers.  Given the quasi-exponential nature of the adoption process these identification methods become impractical for longer times. This is the principal reason why we have not extended the study to later years.  Additionally we have detailed historic accounts of the spreading process  spanning these initial periods only  \cite{DKDTA,KIH}. Such knowledge will allow us to build models below that reflect fundamental social dynamics, different qualitatively from those underlying standard models of epidemics. 
 
\begin{figure} 
\centerline { \includegraphics[width=3.5in,angle=270]{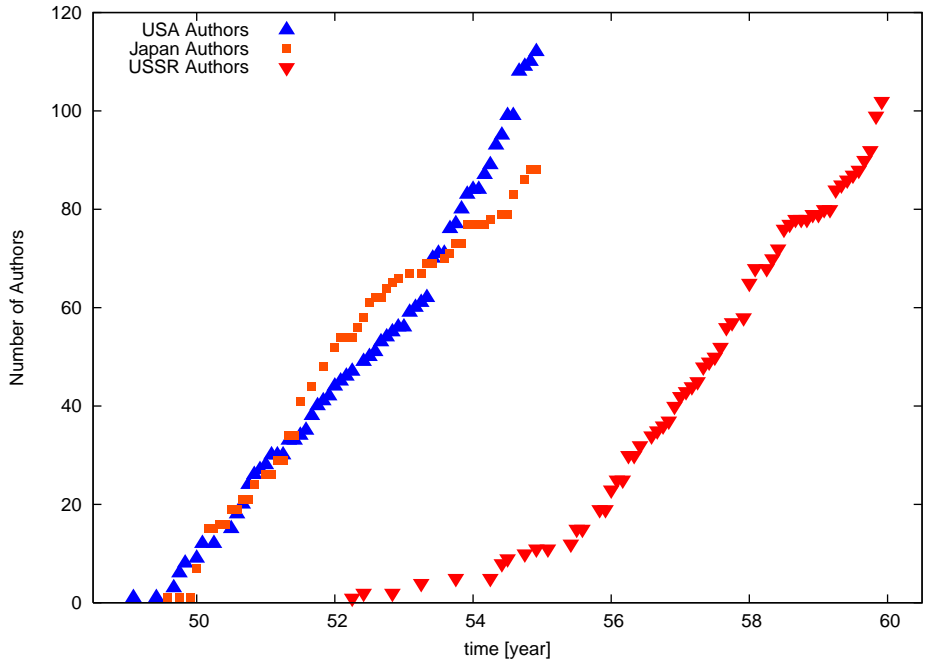} } 
\caption{The time evolution of the cumulative number of authors using
Feynman diagrams in the USA, Japan, and the USSR. The method was first
discovered in the USA and quickly spread both there and in Japan. Adoption
was particularly fast in Japan where researchers had already developed
similar methods.  At the same time, new institutions were developed
throughout Japan after World War II that helped the nation's physicists
share information from the international scientific community that might
otherwise have been difficult to access.  Adoption in the USSR occurred
later because of scientific isolation from physicists in the West with the
onset of the Cold War, and proceeded more slowly because of institutional
resistance.  For details of these institutional and pedagogical factors,
see \cite{DKDTA,KIH}. }
\label{fig1} 
\end{figure}

The identification of adopters with published authors can clearly lead to
underestimation. Similarly the identification of national communities with
specific journal publication is imperfect, although we find almost no
cross-national publications, apart from a few British authors who were in
active contact with developments in the USA and published in the {\it
Physical Review}. As such they are counted as part of the diagram-using
community in the USA.  With these choices the evolution of cumulative
numbers of Feynman-diagram authors is shown in Fig.~\ref{fig1}.

We see that none of the data sets shows saturation in the growth of the adoption of Feynman diagrams.  There are good reasons for this, spanning the initial period covered by the data, shown in  Fig. ~\ref{fig1}, and beyond. The physics graduate student enrollment grew rapidly in the US after World War II, faster than any other field, and was growing especially quickly during the late 1940s when Feynman diagrams were introduced \cite{DKHSPS}. This growth persisted until the late sixties, with an average doubling time of 6.24 years.  Among all subfields of physics particle and nuclear physics, where the diagrams first spread, grew the fastest.  The numbers of new physics grad students in the USSR also increased exponentially, at a rate comparable to that in the US during the postwar period, but quantitative estimates are more uncertain. In Japan,  we know that membership in the new Elementary Particle Theory Group (which consisted largely of interested grad students and postdocs) grew rapidly during this period \cite{DKDTA}.  
Moreover during the mid and late 1950s, the range of applications to which physicists applied Feynman diagrams widened considerably.  No longer the province for nuclear and high-energy physicists alone, many people began to apply the diagrams for problems in solid-state physics and beyond.  This led to another surge in growth of diagram adopters, as new cohorts encountered the diagrams across a growing number of sub-fields of physics.  Compounding this growth, a new generation of textbooks appeared that featured the diagrammatic techniques prominently, ensuring even wider adoption within graduate students' curricula. 

Analogies to other population states commonly used in epidemiological models
are natural but must be properly qualified. The identification of
susceptibles is usually problematic both for diseases and ideas. For
simplicity one may consider the entire population that is not infected (or
recovered), but if the spreading process requires such features as direct
contact with those already infected this may turn out to be a gross
overestimate. With the benefit of hindsight we can see what fraction of
the population actually became infected, but such estimates can clearly
underestimate the class of susceptibles.

Finally it is interesting to discuss the recovered state.  For some
communicable diseases such a state does not exist; as it happens in
HIV and tuberculosis, for which infected individuals remain latent for
extensively long periods. On the other hand, there are infectious diseases
in which an individual acquires immunity right after recovery and will not
get re-infected. This is not true with ideas, a case in which culture is
manifestly different from biology. An idea can recur again and again,
whenever it becomes useful, once it becomes part of an individual's
repertoire.  In many cases (and this is clear in our data for several
authors), an individual might publish in areas where Feynman diagrams are
used, only to later leave the area for good or to return to it later. For
very prolific authors, publication in several areas simultaneously occurs
frequently.

With these caveats in mind we proceed in the next sections to apply
epidemic models to our data. Model parameters will be estimated on the
basis of how well they fit the evolution of adopters.  Furthermore, the
results of these estimates will be subjected to broad bounds imposed by
the solutions' plausibility, given our knowledge of the historical facts.

\section{Population models: drawing parallels between epidemics and idea
diffusion} \label{models}

Below we shall concentrate on the classical, simplest epidemiological models, based on
``homogeneous mixing'' in which state variables are only functions of time.  In a review of epidemiological models Hethcote \cite{het}   introduced their compartmental characterization (e.g. SIR, SIS, SEIR, etc.) within a global analysis of the field.  Such survey also discusses how more complex models can be used to asses the impact of population structure (age, risk, gender, etc.), epidemiological variability (age of infection, variable infectivity, distributed incubation periods, etc.), and scale (spatial, temporal, etc.) on disease dynamics  and control.  Although we have knowledge of some population characteristics (e.g. academic level, institutional location) in our data set we feel it may not be large enough to make such distinctions in a way that will lead to useful quantitative discrimination.

\begin{figure}
\centerline {
\includegraphics[width=5in]{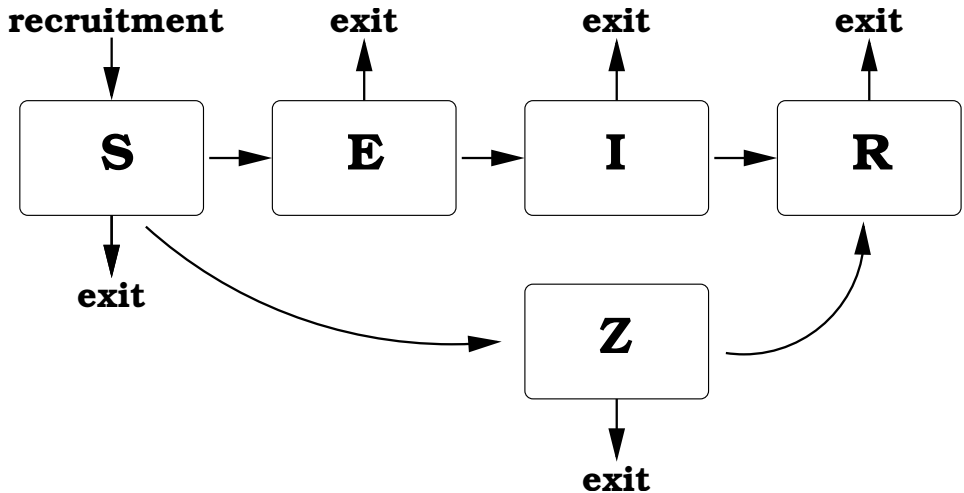}
}

\caption{The basic scheme of population dynamics models for the spread of
ideas, inspired by similar models in epidemics. An individual can be
recruited into the susceptible ($S$) class, then be exposed ($E$) to the
idea, incubate it, and eventually manifest it, becoming a member of the
adopter or infected class ($I$).  An indivdual might instead move into a
competing infective class (e.g., skeptics, $Z$). It is possible that part
of the population may eventually recover ($R$), meaning that it will not
manifest the idea again.  Individuals can also exit any class, thus
reducing the total population.}
\label{fig_epi}
\end{figure}

As such we explore below a large class of "mean-field" models, illustrated by
Fig.~\ref{fig_epi}. At the onset of the spread of the idea most of the
population will be in the susceptible class ($S$), with a few individuals
in the incubator class ($E$) -- having been in contact with the idea --
and a small number of adopters ($I$) manifesting it. These are the
principal classes in the models below.  In addition, inspired by the approaches
of Daley and Kendall \cite{dake1}, we also explore models in
which there may be competing and mutually exclusive ideas (e.g. where
susceptibles are turned off from the idea and become skeptics or idea
stiflers, represented by the class $Z$).  Furthermore, individuals may
recover or become immune ($R$), and not manifest the idea again. Different
models combine subsets of these states and admit different couplings
between them.

The total population is denoted by $N(t)$, where $N=S+E+I+Z+R$.  In the
epidemic models used in this study, the demographic dynamics are modeled by
$d N/dt=\mathcal{B}(N) - \mu N$, where $\mathcal{B}(N)$ is referred to
as the recruitment function.  In our case, this denotes the arrival rate
of new individuals susceptible to the idea, such as new graduate students
starting in the field as well as other scientists who find the idea
relevant for their research.  The parameter $\mu>0$ denotes the rate at
which physicists stop using Feynman diagrams (the exit terms in
Fig.~\ref{fig_epi}). Thus, the maximum value that $1/\mu$ can take is the
average lifespan of the idea within a generation of researchers in the relevant community.
			
	\begin{table}
	\begin{center}	
	\begin{tabular}{|c|l|}\hline
		Variable & Definition \\ \hline
		$S $& Susceptible \\ \hline
		$E $& Idea Incubators \\ \hline
		$I $ & Idea Adopters \\ \hline
		$Z $ & Skeptics \\ \hline
		$R $ & Recovered   \\ \hline
		$N $& Total Population: $N =S +E +I +Z +R $ \\ \hline
	\end{tabular}
	\end{center}
         \caption{Nomenclature for the state variables of the several 
population models used to describe the spread of ideas.}
	\label{vars}
	\end{table}

Whenever $\mathcal{B}(N)>0$ and $\mu >0$, then the system in
Fig.~\ref{fig_epi} is said to have {\it vital dynamics}.  If
$\mathcal{B}(N)\equiv \Lambda>0$, then $N(t)$ varies over time and
approaches a stable fixed point, $\Lambda / \mu$, as $t\rightarrow\infty$, in other words,
the community approaches its ``carrying'' capacity.
In order to illustrate generic model features we discuss below a few
partial implementations of this general scheme, including explicit
parameterizations.  At the end of this section we emphasize the role of
the basic reproductive number, $R_0$, as a measure of effectiveness of
adoption.

\subsection{Models without incubation: SIR Model}

The classical epidemic model consists of three states: susceptibles ($S$),
adopters (or infected, $I$), and recovered ($R$).  In this SIR model,
susceptible individuals transit directly to the adopter class through
contact with other adopters, without any delay period or incubation.  The
recovered state consists of those individuals who no longer manifest the
idea. This state allows for the decay of adopters by recovery and thus
leads to a regulation of the idea spread.  The model is defined by the
following system of ordinary differential equations (where overdots denote
derivatives with respect to time):

\begin{equation}\label{sir}
\left\{
\begin{array}{ll}
\dot S=& \Lambda-\beta S\frac{I}{N}-\mu S,\\\\
\dot I=&\beta S\frac{I}{N}-(\gamma+\mu) I,\\\\
\dot R=&\gamma I-\mu R,\\\\
\dot N=&\Lambda-\mu N ,
\end{array}
\right.  
\end{equation}  
where $1 / (\gamma+\mu)$ is the average time spent manifesting the idea as
an adopter ($\gamma$ denotes the recovery rate from
infection).  The term $\beta S I / N$ is usually referred to as
the standard incidence. The parameter $ \beta $ is the per capita idea
adoption rate. It can in turn be thought as the product between the
mean contact rate per capita and the probability of adoption per
contact.

As noted above, although recovery is a natural concept in epidemiology
(since organisms naturally may become immune after exposure and/or
infection), there is no strict parallel when discussing ideas. Loose
analogies are possible, e.g. once one loses interest in an idea it is
usually harder to have an individual express it, whereas novelty may make
it more attractive. Nevertheless there is no systematic cognitive process,
analogous to the immune system, that actively clears out ideas. As such
many ideas are remembered for life.

Many ideas may be short-lived, say from years to days, compared to the
lifetime of the individual.  In this case, we may consider a single
outbreak by setting $\Lambda=\mu=0$.  The sign of the right hand side of
the second equation in system (\ref{sir}) then determines the spread of
the idea and depends on the initial fraction of susceptibles, $S(t_0) /
N$.  If the initial state of the population can be such that $S(t_0) / N <
\gamma / \beta$, then the number of infectives can only decrease.  This is
the basis of immunization campaigns, whereby members of the susceptible
class are turned into members of the immune class, and hence become part
of $R(t_0)$.  Thus knowledge of the infection rate, $\beta$, and of the
lifetime of the infection, $1 / \gamma$, results in the recommendation for
the fraction of immune (recovered) necessary for an epidemic not to
develop, namely $R(t_0) / N > 1 - (\gamma / \beta)$.  For a very
infectious disease or idea (large $\beta$)  or one with a slow recovery
rate (small $\gamma$) almost all of the population must be immune in order
to halt the spread.
    
Due to the less clear definition of immunity to an idea, the concept of
what may constitute immunization is also ill-defined. Clearly the novelty
of an idea and a perception of its potential are often its most attractive
features. Changing this perception through education (e.g., about the
consequences of a certain behavior, ideology, or practice) may lead to an
increase of skepticism and consequently greater ``immunity'' upon
exposure. Moreover we should keep in mind that this concept of
immunization, just as in standard epidemics but for different reasons, is
usually only valid for the lifetime of an individual. Although some
biological immunity can be passed e.g. from mother to infant, it is
usually the case that young individuals are more susceptible to new
diseases and ideas alike. In the Feynman diagram case this is borne out
historically:  over 80 percent of the early adopters of the diagrams in
each country were either graduate students or postdocs when they first
began using the diagrams; older physicists simply did not re-tool
\cite{DKDTA}.

The asymptotic late-time dynamics of model (\ref{sir}) are well known, and
will form the basis for the analyses of more complex models discussed
below. Suppose that $\Lambda>0$ and $\mu>0$.  For long times, and
regardless of the distribution of infectives and susceptibles,
recruitment and exits will balance each other so that
$\lim_{t\rightarrow\infty}N(t)=N^*=\Lambda / \mu$.  There are up to two
different non-negative steady states (fixed points), known in epidemiology as the {\em
disease-free equilibrium} with $S^*=N^*=\Lambda / \mu$, $I^*=R^*=0$, and
the {\it endemic state} (whenever $\beta/(\gamma+\mu)>1$) with
\begin{eqnarray}
S^*=\frac{\gamma+\mu}{\beta}N^*, \qquad I^* = \left[ \frac{\mu}{\gamma 
+\mu} - \frac{\mu}{\beta} \right] N^*, \qquad  R^*=\frac{\gamma}{\mu}I^* 
= \left[ \frac{\gamma}{\gamma+\mu} - \frac{\gamma}{\beta} \right] N^* .
\end{eqnarray}
The eigenvalues around the disease-free state equilibrium are
$(-\mu,-\mu,\beta-(\gamma+\mu))$.  Thus it is stable provided that
$\beta<\gamma+\mu$, i.e. if the decay rate (due to exit and recovery) is
larger than the idea adoption rate.  The instability of the disease-free state corresponds to
stability of the endemic state.  The eigenvalues of the linearized system
around the endemic equlibrium are
\begin{eqnarray}
-\mu, \ && - \frac{\beta \mu \pm A}{2(\gamma+\mu)}, 
\end{eqnarray}
where $A=\sqrt{\mu( \beta^2 \mu -4 \beta (\gamma+\mu)^2 +
4(\gamma+\mu)^3)} $. All eigenvalues are negative provided that
$\beta>\gamma+\mu$, guaranteeing the local stability of the endemic 
state.

As a result a transcritcal bifurcation (where the two equilibria exchange
stability) takes place at $R_0 \equiv \beta / (\gamma + \mu) = 1$.  In the
mathematical epidemiology literature the dimensionless quantity $R_0$ is
known as the basic reproductive number.  $R_0$ has an intuitive and useful
interpretation as the average number of secondary cases produced by a
``typical'' infected individual during his/her entire life as infectious, when
introduced in a population of susceptibles (assumed to be at a demographic steady state). We will 
discuss the role of $R_0$ further in Subsection \ref{ssR0}.

\subsection{Competition and incubation: SIZ and SEIZ Models}

In the spread of ideas, but almost never in standard epidemics, the
exposure of individuals to an idea almost invariably leads to both
enthusiasts and skeptics. In the case of Feynman diagrams, skeptics did
indeed emerge.  Julian Schwinger, for example, who developed a
non-diagrammatic method of renormalization, quipped years later that
Feynman diagrams had ``brought computation to the masses'' -- hardly a
good thing, as far as Schwinger was concerned.  Although his graduate
students at Harvard did learn something about the diagrams, they made
little use of them in their dissertations and early articles.  J. R.
Oppenheimer, too, was initially skeptical, and effectively blocked Dyson's
recruitment efforts at the Institute for Advanced Study in Princeton for
several weeks, before Hans Bethe interceded directly on Dyson's behalf.  
In Moscow, meanwhile, the influential Lev Landau made his distaste for
Feynman diagrams clear during the early 1950s, blocking any discussion of
them in his famous seminar (even chastising one young graduate student who
had expressed interest in the diagrams that it would be ``immoral'' to
chase such ``fashions'' as Feynman diagrams!) \cite{DKDTA}.  Thus
inclusion of skeptics alongside enthusiasts is quite important.  This can
be modeled by considering two competing and mutually exclusive infected
states, say $I$ and $Z$. The simplest such model (SIZ) is given by
\begin{equation}\label{sizsys}
\left\{
\begin{array}{ll}
\dot S=&\Lambda -\beta S\frac{I}{N} - b S \frac{Z}{N}-\mu S\\\\
\dot I=&\beta S\frac{I}{N}-\mu I\\\\
\dot Z =& b S \frac{Z}{N} - \mu Z ,
\end{array}
\right.
\end{equation}
where $b$ and $\beta$ denote the per capita rates of idea rejection and
adoption by susceptibles, respectively.

The interesting new feature about this type of model is that it can support up to three
fixed points.  The first is the usual disease-free state
$S=N^*= \Lambda / \mu, \ I=0, \ Z=0$ (extinction of both adopters 
and skeptics), and two endemic states, one for each strand $I$, $Z$:
\begin{equation}
\begin{array}{lr}
S=\frac{\mu}{\beta} N^*, \ I=\left( 1 - \frac{\mu}{\beta} \right) N^*, \ 
Z=0& (\mbox{extinction of skeptics})
\end{array}
\end{equation}
or 
\begin{equation}
\begin{array}{lr}
S=\frac{\mu}{b} N^*, \ Z=\left( 1 - \frac{\mu}{b} \right) N^*, \ I=0&  
(\mbox{extinction of adopters}) .
\end{array}
\end{equation}
Observe that model (\ref{sizsys}) does not support the steady state
co-existence of adopters and skeptics.  For the disease-free state the
eigenvalues are $(b-\mu, \beta-\mu,-\mu )$.  Thus for stability one needs
both $b<\mu$ and $\beta<\mu$. This means that there are two
$R_0$'s, $R_0^I=\beta/\mu$, and $R_0^Z = b/\mu$.

Under these circumstances, which of the endemic states becomes stable? To
investigate this question we inspect the eigenvalues around the $I$
endemic state. This gives
 \begin{eqnarray}
 -\mu, \qquad  \left( \frac{b}{\beta} -1 \right) \mu, \qquad  -\beta + 
\mu.
 \label{eigenI}
 \end{eqnarray}
Similarly we obtain the eigenvalues for the endemic $Z$ state by replacing
$b$ with $\beta$ and vice versa in (\ref{eigenI}). This result implies
that only one of the two endemic states can be stable, depending on the
relative magnitude of the contact rates $b$ and $\beta$. We note, however,
that because there is no contact term between the $I$ and $Z$, the way
one class ends up dominating relies on long-time changes in the population
through cycles of recruitment and exit.  This time scale can be very long,
diverging in the limit where $b\rightarrow \beta$.  For $\beta>b$ it will
take on average $\beta/(\beta-b)$ generations until the disappearance of
skeptics.

The model generalizes immediately to an arbitrary number, $n_Z$, of
alternative endemic states, $Z_i$ (in which we include the usual $I$),
with associated contact rates $b_i$. There will then be $n_Z+1$ fixed
points, one disease-free and $n_Z$ endemic corresponding to each strand.
As in the SIZ model above only the state with the largest $b_i$ will be
locally stable.  The stability of the fixed point associated with $Z_i$
for decay in favor of an alternative state $Z_j$ is characterized by an
eigenvalue $[(b_i/b_j) - 1]\mu$.  The disease-free equilibrium will be
locally stable if and only if all $R_0^i = b_i/b_j <1, \
\forall_{i=1}^{n_Z}$.
 
As above, consider the case in which recovery can take place in the SIZ
model, and proceeds with rate $\gamma_I$ from the $I$ class, and with rate
$\gamma_Z$ from the $Z$ class. $R_0^{I,Z}$ change by the simple
modification $\mu \rightarrow \mu+\gamma_{I,Z}$. In the absence of vital
dynamics, it then becomes a necessary and sufficient condition for the
growth of the strand $I$, $Z$ that $S(t_0)/N>\gamma_{I,Z}/\beta$,
respectively. What is interesting now is that the reduction of the
susceptibles can be achieved by having a suitably large fraction of the
population in the complementary infective strand(s). For example, $I$ will
not grow if $[Z(t_0) + R(t_0)]/N > 1 - (\gamma_I/\beta)$. This observation
quantifies the fact that in a population with a large fraction of skeptics
an idea will not take hold. In this sense complementary strands
effectively act like recovery states. This may be the most natural
explanation for why old ideas seldom re-surface, in spite of being
preserved for very long times in the population and various archives.
 
 	\begin{table}
	\begin{center}
	 \begin{tabular}{|c|c|c|}\hline
 	Parameter & Definition\\ \hline 
	$\Lambda$ &  Recruitment rate \\ \hline
	$1/\mu$ &  Average lifetime of the idea   \\ \hline
	$1/\epsilon$& Average idea incubation time  \\ \hline
	$1/\gamma$ & Average recovery time \\ \hline
	$\beta$& Per-capita $S$-$I$ contact rate   \\ \hline
	$\rho$ & Per-capita $E$-$I$ contact rate   \\ \hline
	$b$ & Per-capita $S$-$Z$ contact rate  \\ \hline
	$l$ &    $S\rightarrow Z$ transition probability given contact 
with skeptics \\ \hline
	$1-l$&  $S \rightarrow E$ transition probability given contact 
with skeptics  \\ \hline
	$p$ &  $S\rightarrow I$ transition probability given contact 
with adopters \\ \hline
	$1-p$& $S\rightarrow E$ transition probability given contact 
with adopters \\ \hline
 \end{tabular} 
\caption{Parameter definitions used in the several population models of 
this section.}      
\end{center}  
\end{table}

One important drawback of SIR and SIZ models is that once exposed to an
infected person, a susceptible individual transits immediately to the
infected class.  This feature is often unrealistic, especially for ideas
that require long periods of apprenticeship, which is common in scientific
research and is a significant feature of the Feynman-diagram user data
discussed below.  The simplest way of incorporating some delay in an SIZ
model is to introduce a new class of incubators (or exposed), denoted by
$E$, between the susceptible and adopter states. Upon contact with an
adopter, a susceptible individual transits with a given probability to the
$E$ class.  This class has a given characteristic lifetime, $1 /
\epsilon$, before the individual manifests the idea and transits to the
$I$ class. That is, $1 / \epsilon$ is the average incubation (or
maturation) time of the idea \cite{incub}. It is expected to be a function
of personal effort on the part of the adopter as well as environment
(adverse or supportive). There may also be population losses due to vital
dynamics, which we will continue to assume occur on a timescale $1 / \mu$.
In this sense not all of the exposed population will become infected.

This extension leads to an SEIZ model.  In addition, this model can be
enriched with extra processes to generate a better description of the
data.  Below we present a version of the SEIZ model in which skeptics
recruit from the susceptible pool with rate $b$, but their action may
result either in turning the individual into another skeptic (with
probability $l$), or it may have the unintended effect of sending that
person into the incubator class (with probability $1-l$). We also
introduce a probability, $p$, that a susceptible individual will become
immediately infected with the idea upon contact.  Conversely, with
probability $1-p$ that person will transit to the incubator class instead,
from which the individual may later become an adopter. Furthermore, the
transition of individuals from the incubator class to the adopter class
can be promoted by contact, with rate $\rho$.  With these choices the
model is given by:

\begin{equation}\label{seiz}
 \left\{
\begin{array}{ll}
\dot S=& \Lambda-\beta S\frac{I}{N}-b S\frac{Z}{N} -\mu S\\\\
\dot E=&(1-p)\beta S\frac{I}{N}+(1-l) b S\frac{Z}{N}-\rho 
E\frac{I}{N}-\epsilon E-\mu E\\\\
\dot I=&p\beta S\frac{I}{N}+\rho E\frac{I}{N}+\epsilon E-\mu I\\\\
\dot Z=&l~ b ~S\frac{Z}{N}-\mu Z .
\end{array}
\right.
\label{SEIZ}
\end{equation}
As expected, the system has a disease-free state with $S^*=N, \
E^*=I^*=Z^*=0$. Analysis of the local stability of this fixed point 
(utilizing next generation operator \cite{casfenhua1,vadwa1}) 
reveals that the basic reproductive numbers are given by
\begin{equation}
R_0^{I,Z}=\left(  \frac{\beta(\epsilon + p\mu)}{\mu(\epsilon+\mu)}, 
\frac{l ~b}{\mu}  \right) .
\label{R0IZ}
\end{equation}
As in the SIZ model the first number, $R_0^I$, is the one of interest, as
it corresponds to an eigenvector with a component of adopters. The second
value, $R_0^Z$, corresponds to the exclusive growth of a population of
skeptics, without acceptance of the idea.

\subsection{Speed of idea propagation and effectiveness of adoption}
\label{ssR0}

From the discussion of the models above we can define several important
intuitive quantities that characterize the spread of ideas. For example, a
simple measure of the speed of propagation of the idea is the number of
new adopters per unit time. This is simply given by $\dot I$.
 
For simple models, such as the ones discussed above, in which there is 
only
one growing eigenvalue $\lambda^+$ for each infective strand, the initial
velocity of the spread is simply
\begin{eqnarray}
v_{\rm ini}\equiv \dot I (t_0)\simeq \lambda^+ I(t_0).
\end{eqnarray}
The quantity $v_{\rm ini}$ gives a measure of how fast the idea will
initially spread but not of its overall adoption effectiveness. In order
to determine the latter we must consider the number of new adoptions that
a spreader of the idea can lead to during his/her lifetime. Since there is
no {\it a priori} good reason to suspect that ideas are short-lived, the
effectiveness of an idea may result from slow spread over long periods of
time and thus may not be well characterized by $v_{\rm ini}$.

The number of secondary adoptions induced by a typical idea spreader in a
population of susceptibles over that person's lifetime as an adopter,
$t_{\rm idea}$, is called the {\it basic reproductive number}, $R_0$, in
ecology and epidemiology (see \cite{thi, dihe, het}).  As such $R_0$ is
the invasion criterion for adopters in a population of susceptibles, or
analogously the average branching ratio (the number of offspring) of the
typical adopter over his/her lifetime in this state. If $R_0=I(t_{\rm
idea}) / I(t_0) > 1$ then the idea will spread. The greater $R_0$ the more
effective the idea adoption will be.

In practice $R_0$ can be computed in simple models through the
linearization of $\dot I(t)$ around the disease-free equilibrium. These
expressions are summarized in Table \ref{r0s}. For the computation of
$R_0$ in models with heterogeneous populations other methods are
necessary \cite{dihe,casfenhua1,vadwa1}.  In the next section we will
estimate the statistical distributions for $R_0$ subject to fitting the
data for the spread of Feynman diagrams in the USA, Japan, and the USSR.  
The mean of this distribution provides a measure of the effectiveness of
the adoption of Feynman diagrams in the three countries.

\begin{table}
	\begin{center}
		\begin{tabular}{|c|c|c|c|}\hline
		Model &~ SIR ~& SEI& SEIZ \\ \hline
		& && \\
		 $R_0^I$& $\frac{\beta}{\gamma+\mu}$ 
		 &$\frac{\beta \epsilon}{\mu (\mu +\epsilon)}$ 
		 &$\frac{\beta(\epsilon+p\mu)}{\mu(\epsilon+\mu)}$  \\
		&&& \\ \hline
		\end{tabular}
	\end{center}
	\caption{Basic reproductive number $R_0^I$ for the 
SIR, SEI, and SEIZ models discussed in section \ref{models}.}
	\label{r0s}
\end{table}

\section{Results and discussion}       
\label{results}    
        
We now analyze the results of estimating parameters by matching the data
on the spread of Feynman diagrams for three distinct countries to
several population models discussed above.  These models allow us to
discuss the effects of the recovered class, of latency, and of competitive
idea strands. They also explore several classes of transition mechanisms,
both by progression and by contact between population classes.
       
\begin{table}
\begin{center}
\begin{tabular}{|c|c|c|c|} \hline
	model         &  USA      & Japan   &  USSR    \\ \hline	
         	SIR      &  2.816     &  1.788   &   1.487  \\ \hline
         SEI       &  1.963     &  1.638   &   1.437  \\ \hline 
	SEIZ            & 1.467      &  1.568   &   1.437  \\ \hline
	\end{tabular}
\end{center}
\caption{The smallest (absolute value) average deviation per data point
between the best fit parameters of each model and data on the number of
Feynman diagram adopters for the USA, Japan, and the USSR. }
\label{tabdev}
\end{table}

Table \ref{tabdev} summarizes the results. To gauge the applicability of
each model to each data set we used the simplest measure of goodness of
fit, by computing the absolute value of the deviation between model
prediction and data. Average deviations per data point are shown in Table
\ref{tabdev}. Details of our ensemble estimation procedure are given in
Appendix A. 

Here we note simply that parameter estimation must, by practical
necessity, be confined to given numerical ranges, with upper and lower
bounds dictated by general empirical considerations. Our choices of
estimation intervals are shown in Table \ref{tabpara}. This procedure is
familiar from epidemiology, where knowledge about such quantities as the
length of incubation and infectious periods is often used to restrict
various model parameters to plausible values (see \cite{chfecaca};
\cite{penemaleho} also employs assumptions of this nature in immunology).

\begin{table}
\begin{center}
 \begin{tabular}{|c|c|c|}\hline
 parameter & baseline range & unit \\ \hline
 $S(t_0)$ & [0,500] &  people   \\ \hline
 $E(t_0)$ & [0,100] & people \\ \hline
 $I(t_0)$ & [0,20] &   people  \\ \hline
 $R(t_0)$ & [0,10]  &  people  \\ \hline
 $Z(t_0)$ & [0,100] & people \\ \hline
 $\epsilon$ & [0.2,6] & 1/year \\ \hline
 $\beta$ & [0,12] &  1/year   \\ \hline
 $b$ & [0,12] &  1/year   \\ \hline
 $l$ & [0,1] & 1 \\ \hline
 $\gamma$ & [0,12]  &  1/year  \\ \hline
 $\Lambda$ & [0,50] &  people/year  \\ \hline
 $\mu$ & [0.025,12] &   1/year  \\ \hline
 $p$ & [0,1] & 1 \\ \hline
 $\rho$ & [0,12]  & 1/year \\ \hline
 \end{tabular}
\end{center}
\caption{
Parameters used in the SIR, SEI and SEIZ population models, 
their allowed ranges in our estimation procedure and units.}
\label{tabpara}
\end{table}

\subsection{Results for models without incubation: SIR}

\begin{figure}
\centerline{
\includegraphics[width=3.5in,angle=270]{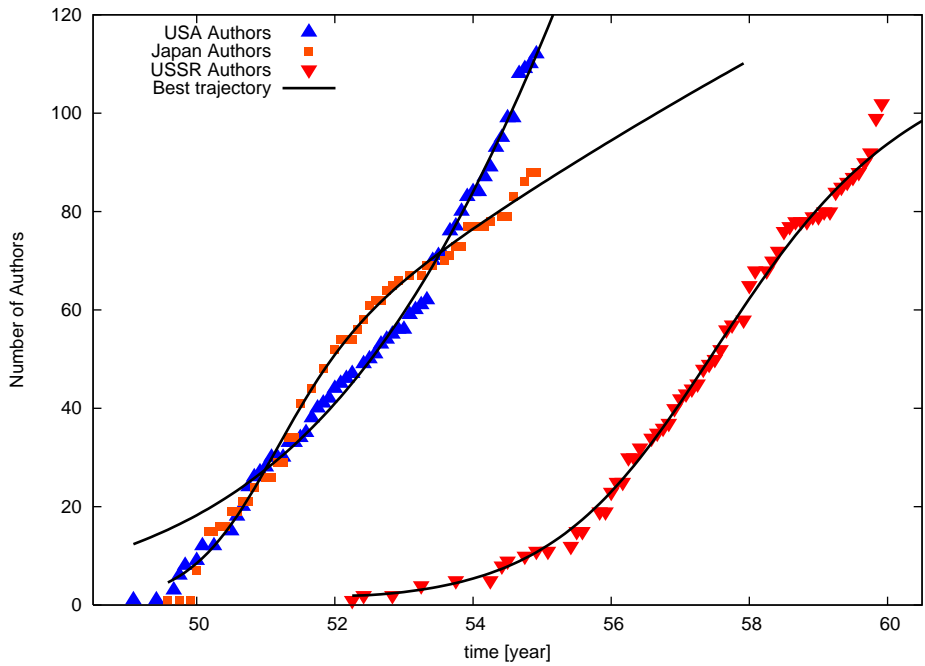}
}
\caption{The best fit trajectory (see  table \ref{tabSIR}) for the rise 
of Feynman diagram adoption obtained for the SIR model vs. the data for 
the USA, Japan, and the USSR. 
}
\label{figSIR}
\end{figure}

We start by presenting our results for simple models without incubation.
Parameter estimates are given in Table~\ref{tabSIR} for the USA, Japan,
and the USSR, while the model solutions are compared to the data in
Fig.~\ref{figSIR}.

\begin{table}
	\begin{center}
	\begin{tabular}{||c|c|c|c||}\hline
		\multicolumn{4}{|c|}{\bf USA} \\ \hline\hline
		\multicolumn{1}{|c|}{parameter}
		&\multicolumn{1}{|c|}{best-fit}
		&\multicolumn{1}{|c|}{mean}
		&\multicolumn{1}{|c|}{std}\\ \hline
		$S(t_0)$ &   114.092  &   96.463  &   76.726  \\ \hline
 		$I(t_0)$ &   11.948 &   10.982  &   0.542  \\ \hline
 		$R(t_0)$ &   0.830  &   0.550  &   0.432  \\ \hline
 		$\beta$ &   0.534  &   0.663  &   0.052  \\ \hline
 		$\gamma$ &  $8.542\times 10^{-3}$ &   0.049  &   0.034  \\ \hline
 		$\Lambda$  &   40.417  &   42.864  &   5.130  \\ \hline
		$\mu$  &   0.036 &   0.058  &   0.023  \\ \hline
		$R_0$ &   12.029  &   6.752  &   2.008  \\ \hline
	\end{tabular}
	\begin{tabular}{||c|c|c|c||}\hline
		\multicolumn{4}{|c|}{\bf Japan} \\ \hline\hline
		\multicolumn{1}{|c|}{parameter}
		&\multicolumn{1}{|c|}{best-fit}
		&\multicolumn{1}{|c|}{mean}
		&\multicolumn{1}{|c|}{std}\\ \hline
		 $S(t_0)$  &   33.901 &   24.534  &   3.537  \\ \hline
	 	 $I(t_0)$ &   4.018  &   3.799 &   0.348  \\ \hline
		$R(t_0)$   &   1.925 &   0.864  &   0.714  \\ \hline
		$\beta$  &   1.990 &   2.255  &   0.131  \\ \hline
		$\gamma$  & $8.668\times 10^{-3}$  &   0.054  &   0.034  \\ \hline
		$\Lambda$  &   12.466  &   20.759 &   2.646 \\ \hline
		$\mu$  &   0.031  &   0.087  &   0.037  \\ \hline
		$R_0$ &   49.582  &   16.922  &   4.308  \\ \hline
	\end{tabular}

	\begin{tabular}{||c|c|c|c||}\hline
		\multicolumn{4}{|c|}{\bf USSR} \\ \hline\hline
		\multicolumn{1}{|c|}{parameter}
		&\multicolumn{1}{|c|}{best-fit}
		&\multicolumn{1}{|c|}{mean}
		&\multicolumn{1}{|c|}{std}\\ \hline
		$S(t_0)$   &   1.347  &   1.156  &   1.088  \\ \hline
 		$I(t_0)$  &   1.935  &   1.583 &   0.218  \\ \hline
		$R(t_0)$  &   9.742  &   4.928 &   2.415  \\ \hline
		$\beta$  &   1.251 &   1.258  &   0.045  \\ \hline
		$\gamma$ &   0.030  &   0.092  &   0.062  \\ \hline
		$\Lambda$   &   32.822 &   32.031  &   6.894  \\ \hline
		$\mu$ &   0.188 &   0.134  &   0.063  \\ \hline
		$R_0$ &   5.739 &   6.053  &   1.963  \\ \hline
	\end{tabular}	
	\end{center}
	\caption{Parameter estimation (SIR model) for the spread of
Feynman diagrams in the USA, Japan, and the USSR. The three columns show
our best-fit estimate, the mean computed over an ensemble of parameter set
realizations, and corresponding standard deviation (std).} 
\label{tabSIR}
\end{table}

\begin{figure}
\centerline{
\includegraphics[width=3.5in,angle=270]{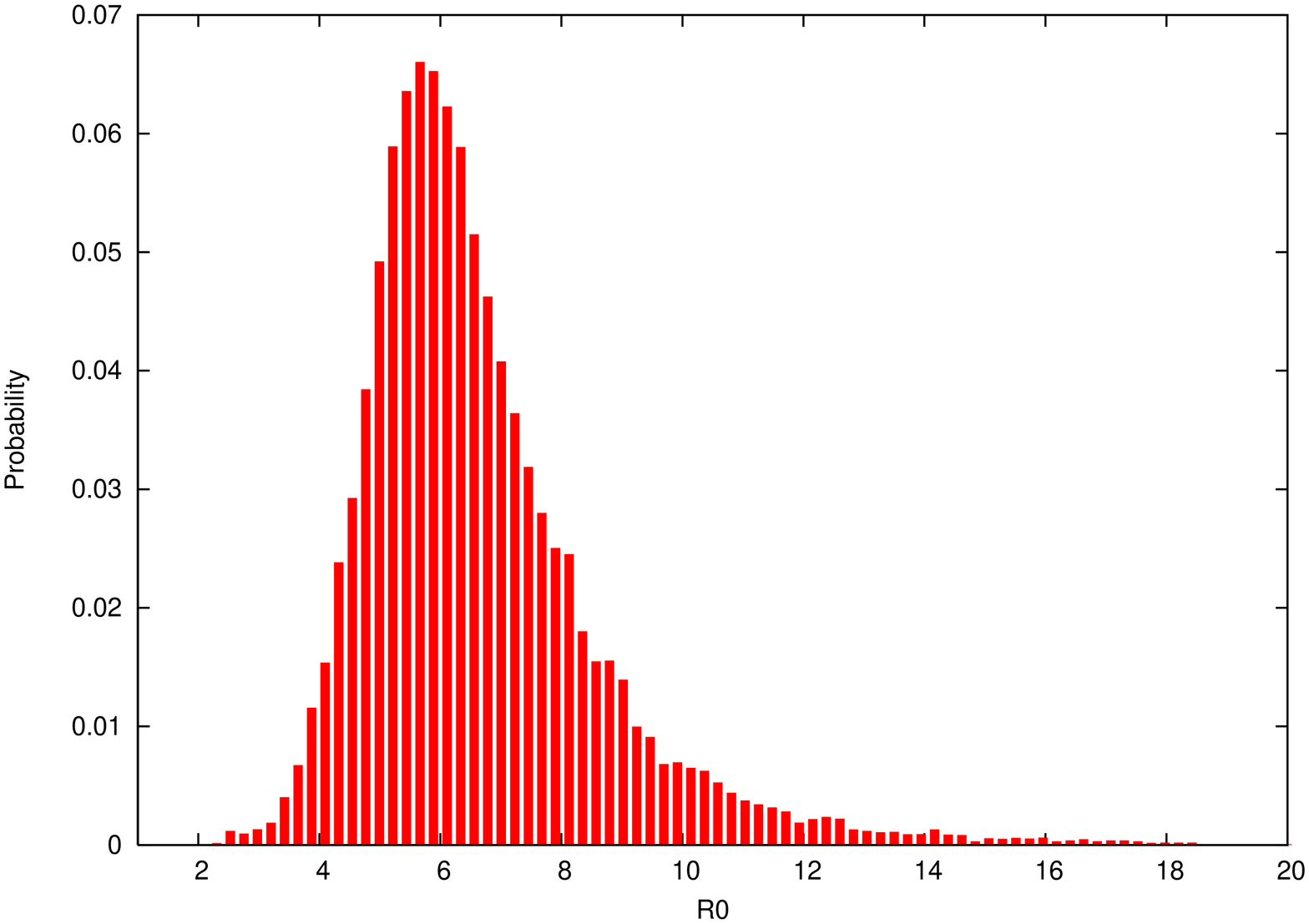}
}
\caption{The probability distribution of the basic reproductive number 
$R_0$ estimated from the SIR model for the USA data on the spread of 
Feynman diagrams. $R_0$ measures the effectiveness of the idea adoption. 
Its value estimates the average number of adopters induced by a typical 
spreader in a population of susceptibles.}
\label{figR0_SIR}
\end{figure}

The estimates for the initial population paint a picture of a considerably
larger scientific community susceptible to learn Feynman diagrams in the
USA than in the other two countries.  In Japan, $S(t_0)$ appears more than
three times smaller than in the USA, while in the Soviet Union our
estimates indicate a very small number of susceptibles around 1952.
Nevertheless both the USA and the USSR show strong levels of recruitment
(slightly higher $\Lambda$ in the USA), as compared to Japan.

This makes sense given each community's rates of growth during this time
period.  In the postwar United States, the rate at which new Ph.D.s in
physics were granted grew by nearly twice that of any other field between
1945 and 1951, quickly exceeding (by a factor of three) the prewar rate at
which new physicists had been trained.  Meanwhile, building on the wartime
Manhattan Project pattern, the federal government pumped money into
physics at more than ten times the prewar levels.  Most singled out for
support during the early postwar period was high-energy physics, precisely
that branch of the discipline in which Feynman diagrams were first
developed and from which the earliest adopters came \cite{DKHSPS}. These
factors led to a large population of susceptibles when Feynman and Dyson
first introduced Feynman diagrams.

Japan, on the other hand, had a strong tradition of high-energy physics
before the war, but massive shortages of funding and basic supplies during
the early postwar years hampered the growth of that country's physics
community (lower $\Lambda$). Although absolute numbers of new physicists
in Japan did not grow at anything like the pace in the United States after
World War II, several institutional changes were introduced in Japan right
around the time that Feynman diagrams were invented, greatly facilitating
the diagrams' spread throughout Japan.  This fact is reflected in the
highest adoption rate, $\beta$, for Japan, compared to the other two
countries. This in turn leads both to the fastest speed of adoption and
the highest value of $R_0$.

Contacts between Japanese and American physicists began again in 1948
(while Japan was still under U.S. occupation), including visits by several
Japanese theoretical physicists to the Institute for Advanced Study in
Princeton, New Jersey, where Freeman Dyson was honing the new diagrammatic
techniques.  A new organization in Japan, known as the Elementary Particle
Theory Group, was also founded in 1948, and began to publish its own
informal newsletter and preprint organ, {\it Soryushi-ron Kenkyu}, which
helped to spread news of the new diagrammatic techniques.  And finally the
Japanese university system quickly expanded tenfold, beginning in 1949,
allowing young physicists to establish new groups and visit new
institutions throughout the country, putting the new techniques into rapid
circulation \cite{DKDTA,KIH}.

The Soviet Union was the only country in the world after World War II in
which the growth in the numbers of new physicists and in government
spending on physics was comparable with the United States.  This may
explain why our estimates of the recruitment rates $\Lambda$ are so high
and commensurate for the two nations. But the onset of the Cold War in the
late 1940s effectively ended all informal communication between physicists
in the USA and USSR just months before Feynman and Dyson introduced
Feynman diagrams.

These geopolitical constraints severely limited the exchange of
information for several years and explain why Feynman diagrams took hold
in the Soviet Union only later and at a slower initial pace (smallest
$v_{\rm ini}$).  Only with the ``Atoms for Peace'' initiatives, starting
in 1955, did physicists from both countries begin to meet informally for
extended visits.  And only after these lengthy face-to-face ``exposures''
did Soviet physicists begin to adopt Feynman diagrams at a comparable rate
to those in the USA and Japan \cite{DKDTA,KIH}. Over time the
effectiveness of adoption, $R_0$, was nevertheless comparable between the
USSR and the USA.

Finally we notice that both the exit and recovery rates, $\mu$ and
$\gamma$, are small in every case, their sum being comparable to a career
lifetime (5-25 years). The fact that $\gamma$ is estimated to be smaller
that $\mu$ is a consequence of our imposed lower bound on the exit rate
and the fact that the data only constrains their sum. Although this
estimate cannot be made with good confidence for data which only covers
the first six years, it is an indication that ideas are not naturally
forgotten.

Incidentally, we do know in some cases that the time to change research
subject was much shorter for a few prominent authors. Richard Feynman was
working almost exclusively on his theory of superfluidity by 1953
(although some of his students continued to use the diagrams under his
supervision), while Freeman Dyson was persuaded to change research
direction, to condensed-matter theory, at a meeting with Enrico Fermi also
in 1953. (See Dyson's testimony in \cite{DysonNature}; see also
\cite{DKDTA}.)

The long exit and recovery times, combined with finite, plausible values
of the contact rate $\beta$, lead in turn to large values of $R_0$. The
fact that an infected individual, when introduced in a population of
susceptibles, can lead to many adopters (here 6-50) is associated {\it
not} with high adoption rate for the idea, $\beta$, but rather with a long
time (many years) over which the idea can be transmitted,
$1/(\gamma+\mu)$. This is a feature that we will see repeated in more
complex models and that is manifestly different between biological
infection and the spread of ideas.

\subsection{Results for models with Incubation: SEI}

We now analyze the effects of including latency in the models. In the
simplest SEI model, susceptibles transit to an intermediate class of
incubators ($E$) upon contact with adopters, in which they remain for a
characteristic ``incubation'' time $1/\epsilon$, after which they manifest
the idea. Note that due to exit processes the average time spent in the
incubator class is actually $1/(\epsilon+\mu)$, and that some individuals
exit the population and never manifest the idea. In practice $\mu$ will be
estimated to be small and the time spent in the incubators class is indeed
essentially the incubation time. The simplest SEI model is a subset of the
SEIZ model of eq.~(\ref{SEIZ}), and is given by
\begin{equation}\label{sei}
 \left\{
\begin{array}{ll}
\dot S=& \Lambda-\beta S\frac{I}{N} -\mu S\\\\
\dot E=&\beta S\frac{I}{N}-\epsilon E-\mu E\\\\
\dot I=&\epsilon E-\mu I .
\end{array}
\right.
\end{equation}
Results of the parameter estimates are presented in Table ~\ref{tabSEI}.

\begin{table}
	\begin{center}
	\begin{tabular}{||c|c|c|c||}\hline
		\multicolumn{4}{|c|}{\bf USA} \\ \hline\hline
		\multicolumn{1}{|c|}{parameter}
		&\multicolumn{1}{|c|}{best-fit}
		&\multicolumn{1}{|c|}{mean}
		&\multicolumn{1}{|c|}{std}\\ \hline
		 $S(t_0)$ &   478.515 &   398.691 &   61.990  \\ \hline
 		$E(t_0)$ &   60.989 &   44.686 &   4.728  \\ \hline
 		$I(t_0)$ &   0.020 &   0.160  &   0.135  \\ \hline
 		$\epsilon$ &   0.257  &   0.391  &   0.055  \\ \hline
 		$\beta$ &   1.041  &   0.951 &   0.086 \\ \hline
 		$\mu$ &   0.025 &   0.040 &   0.012  \\ \hline
 		$\Lambda$ &   45.385 &   40.052  &   6.467  \\ \hline
 		$R_0$ &   37.711  &   23.172 &   5.798 \\ \hline
	\end{tabular}
	\begin{tabular}{||c|c|c|c||}\hline
		\multicolumn{4}{|c|}{\bf Japan} \\ \hline\hline
		\multicolumn{1}{|c|}{parameter}
		&\multicolumn{1}{|c|}{best-fit}
		&\multicolumn{1}{|c|}{mean}
		&\multicolumn{1}{|c|}{std}\\ \hline
		$S(t_0)$ &   30.248 &   31.037  &   2.190 \\ \hline
 		$E(t_0)$ &   11.569&   12.022 &   1.400  \\ \hline
		$I(t_0)$ &   0.153 &   0.165  &   0.129 \\ \hline
 		$\epsilon$ &   2.361 &   2.009 &   0.279  \\ \hline
 		$\beta$  &   5.956  &   4.417 &   0.787  \\ \hline
 		$\mu$  &   0.039  &   0.044 &   0.013  \\ \hline
  		$\Lambda$&   12.067 &   12.578  &   1.093  \\ \hline
 		$R_0$ &   150.136 &   105.372 &   35.223  \\ \hline
	\end{tabular}

	\begin{tabular}{||c|c|c|c||}\hline
		\multicolumn{4}{|c|}{\bf USSR} \\ \hline\hline
		\multicolumn{1}{|c|}{parameter}
		&\multicolumn{1}{|c|}{best-fit}
		&\multicolumn{1}{|c|}{mean}
		&\multicolumn{1}{|c|}{std}\\ \hline
		$S(t_0)$ &   3.074  &   0.810  &   0.722  \\ \hline
 		$E(t_0)$ &   3.344 &   3.462  &   0.647  \\ \hline
 		$I(t_0)$ &   0.682  &   0.738  &   0.266  \\ \hline
  		$\epsilon$ &   1.713  &   1.613  &   0.476  \\ \hline
 		$\beta$  &   3.715  &   3.589  &   0.753  \\ \hline
  		$\mu$  &   0.067  &   0.075  &   0.035  \\ \hline
 		$\Lambda$ &   17.819 &   19.372  &   3.668  \\ \hline
 		$R_0$ &   53.257  &   55.892  &   28.788  \\ \hline
		\end{tabular}	
	\end{center}
 \caption{Parameter estimations for the SEI model for the adoption of  
Feynman diagrams in the USA, Japan, and the USSR.}
	\label{tabSEI}	
\end{table}

\begin{figure}
\centerline{
\includegraphics[width=3.5in,angle=270]{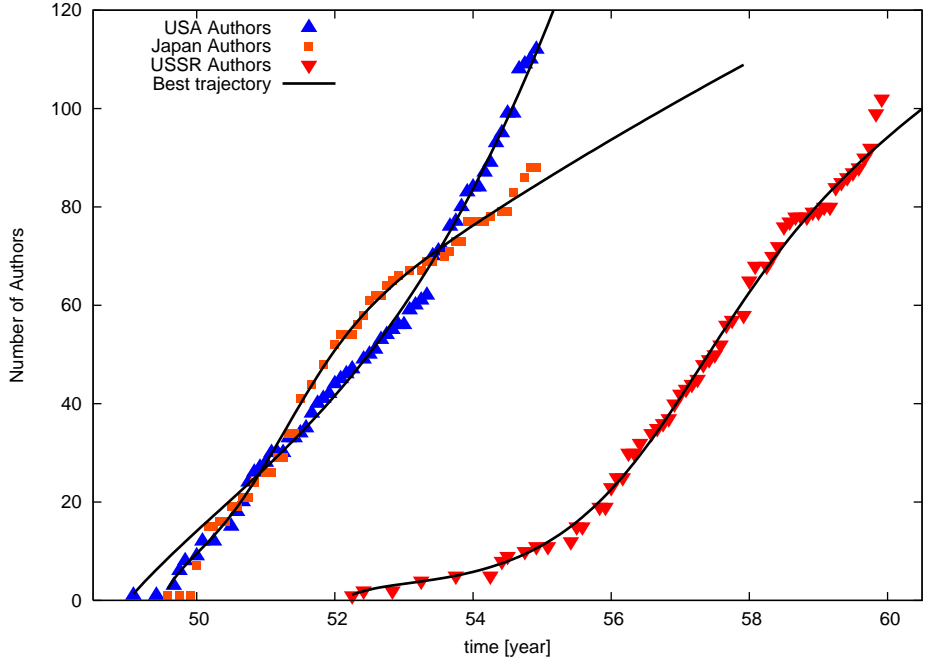}
}
\caption{Best fit trajectories corresponding to the parameter 
estimations for the SEI model, for the USA, Japan, and the USSR. The SEI 
model fits the data better at early times, especially for the USA, when 
compared to the SIR model.}
\label{SEI}
\end{figure}

The most important qualitative difference, relative to models without
latency, is that the model can now better fit data at early times for the
USA and Japan (see Fig.~\ref{SEI}). This accounts for the bulk of the
improvements in Table~\ref{tabdev}.  In both cases this is made
possible in the SEI model because starting with a number of individuals in
the incubator class allows a two-stage growth process for the adopters.  
Initially the incubators are depleted, allowing for a growth of adopters
with a negative second derivative. This is the main feature of SEI
solutions, accounting for their better fit of the data relative to the SIR
model. The two-stage process is a property of the growth curve for
adopters in the USA from the initial time until early 1950, and to a
lesser extent for Japan over the same period, after a slightly later
start.  The characteristic time at which the curve changes concavity can
be computed from the initial growth as
\begin{eqnarray}
t^* \simeq \frac{1}{\epsilon} \ln \left[ \frac{E(t_0)}{E(t_0)+I(t_0)} 
\left( 1+ \frac{\epsilon+\mu}{\beta} \frac{N(t_0)}{S(t_0)} \right) 
\right] .
\end{eqnarray}
This time is longest for the USA, on the order of 10 months, shortest for
Japan at 2.3 months, and about 5 months for the USSR, reflecting the
relative values of the parameters $\epsilon$ and $\beta$ estimated for the
three countries.  Beyond this point in time the effect of the incubator
class is relatively negligible. For Japan and the USSR, where $\epsilon$
is largest, the class becomes essentially non-dynamical beyond the initial
transient, with $E \rightarrow \beta S I/(N \epsilon)$, and as a
consequence the solutions look much like those of the SIR model.

\begin{figure}
\centerline{
\includegraphics[width=3.5in,angle=270]{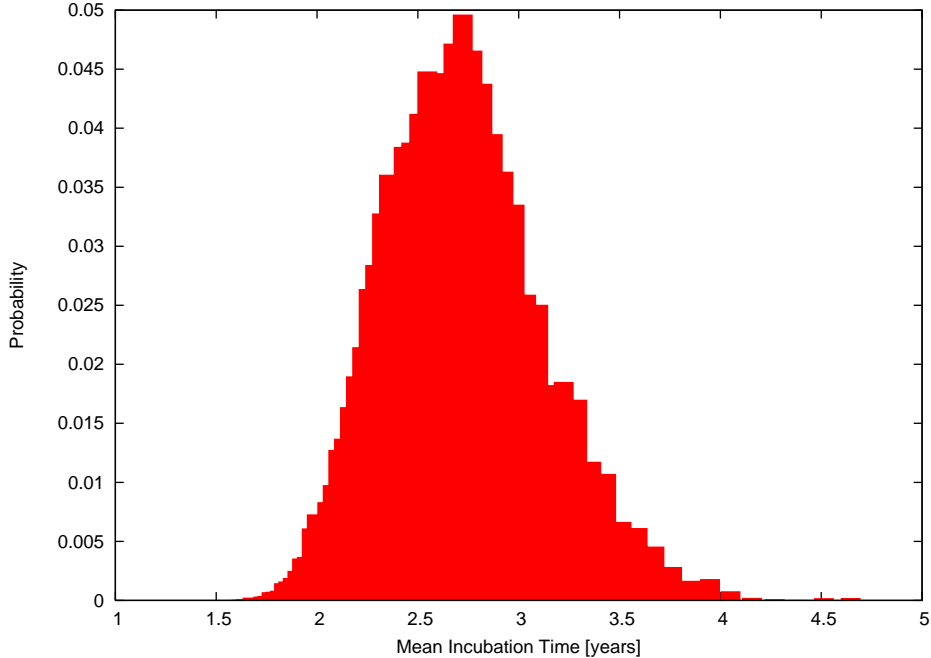}
}
\caption{Histogram of the mean incubation time $1/\epsilon$ for Feynman 
diagrams in the USA, estimated 
by fitting the SEI model to data, see table \ref{tabSEI}.}
\label{figSEI_epsilon}
\end{figure}

In practice the incubation periods estimated for the three countries are
quite different. For the USA (see Fig.~\ref{figSEI_epsilon}), the best fit
solutions prefer to start in 1949 with a relatively large number of
incubators and an incubation time of order 3-4 years.  In both Japan and
the Soviet Union the initial population included fewer incubators but had
a considerably shorter incubation time, of the order of 5-6 months in
Japan and 7-8 months in the USSR.  These incubation period estimates for
Japan and the Soviet Union are unexpectedly short, since most of the
papers were authored by graduate students who took on average a few years
of training (``incubation'') before publishing.  The small values for
$\epsilon$ thus reveal some limits of the simple SEI model:  in
particular, simple progression to adoption (parameterized by $\epsilon$)
does not capture the dynamics adequately, since (as we know historically)
the role of multiple contacts was important.  We return to this issue
below.

Beyond the role played by the incubator class, we observed the same
relative hierarchy of several important quantities among the different
national communities.  Japan had the largest effectiveness of adoption,
$R_0$, whereas both the USA and the USSR displayed smaller and
statistically commensurate values for $R_0$.  (Data were only collected
for the USA and Japan for the period 1949-54, because the steep rate of
growth made longer collection times infeasible.  The slow rise of diagram
adoption in the USSR, on the other hand, encouraged us to collect data for
a longer period, 1949-59, making direct comparisons between late-time
behavior in the USA and the USSR difficult.)  In every case the large
values of $R_0$ are essentially due to a long lifetime of the idea,
$1/\mu$, of 13-40 years.  The recruitment rates, $\Lambda$, similarly to
the SIR estimates, are highest for the USA, followed by the USSR,
reflecting these national efforts to increase the numbers of new
physicists.
   
In spite of all these qualitative similarities one should also keep in
mind that the numerical values for each of these parameters are generally
different between the SIR and SEI models, and not always statistically
compatible. Thus preference of one model over another can be determined
via consideration of the goodness of fit (Table~\ref{tabdev}), but should
also take into consideration qualitative knowledge of the processes at
play.

\subsection{Results for models with Incubation and Competition: SEIZ}

\begin{figure}
\centerline{
\includegraphics[width=3.5in,angle=270]{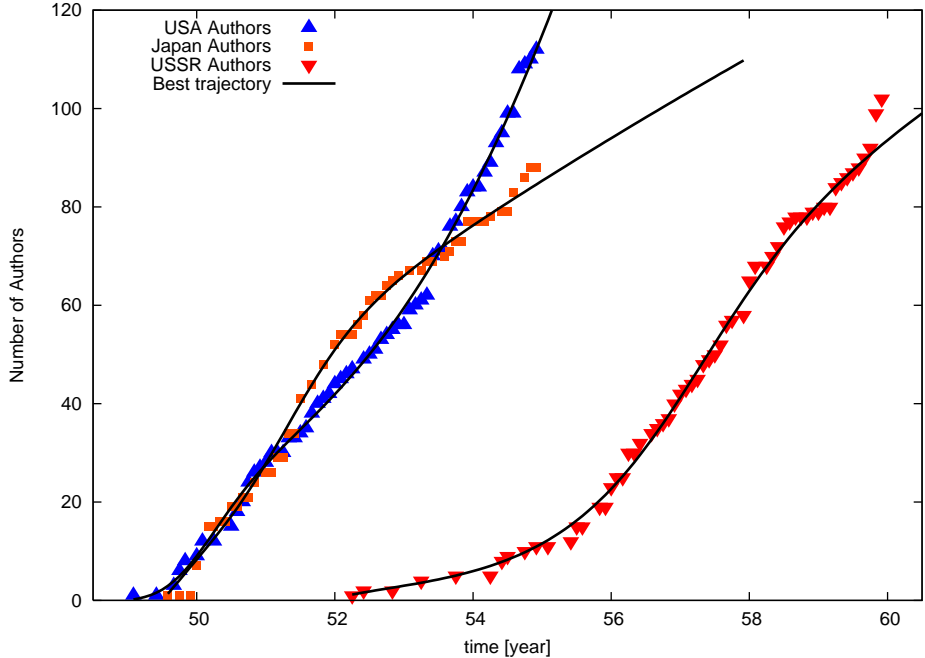}
}
\caption{The best fit solutions of the SEIZ model (see Table 
\ref{tabSEIZ}) vs. the data for the USA, Japan, and the USSR.}
\label{figSEIZ}
\end{figure}

Finally we consider the most complex model of our set, which includes an
additional class $Z$ much like that of adopters, but which competes with
$I$ for susceptibles.  Results of the parameter estimation procedure are 
given in Table~\ref{tabSEIZ} and in Fig.~\ref{figSEIZ}.

\begin{table}
	\begin{center}
	\begin{tabular}{||c|c|c|c||}\hline
		\multicolumn{4}{|c|}{\bf USA} \\ \hline\hline
		\multicolumn{1}{|c|}{parameter}
		&\multicolumn{1}{|c|}{best-fit}
		&\multicolumn{1}{|c|}{mean}
		&\multicolumn{1}{|c|}{std}\\ \hline
		$S(t_0)$ &   98.973 &   108.662 &   5.852 \\ \hline
 		$E(t_0)$ &   24.515 &   24.984 &   0.447  \\ \hline
 		$I(t_0)$ &   $5.916\times 10^{-5}$ &   0.031  &   0.027  \\ \hline
	 	$Z(t_0)$ &   0.114 &   0.160  &   0.119  \\ \hline
 		$\epsilon$ &   0.202 &   0.210  &   0.009  \\ \hline
 		$\beta$ &   0.488  &   0.496  &   0.012  \\ \hline
 		$b$  &   0.164  &   0.156  &   0.117  \\ \hline
 		$l$ &   0.311  &   0.252  &   0.171  \\ \hline
 		$\mu$ &   0.025  &   0.032  &   0.006 \\ \hline
		$p$  &   0.570  &   0.566  &   0.052  \\ \hline
		$\rho$ &   11.893  &   11.549  &   0.330  \\ \hline
 		$\Lambda$  &   49.527  &   47.860 &   1.555  \\ \hline
 		$R_0^I$ &   18.412  &   14.975  &   2.227  \\ \hline
	\end{tabular}
	\begin{tabular}{||c|c|c|c||}\hline
		\multicolumn{4}{|c|}{\bf Japan} \\ \hline\hline
		\multicolumn{1}{|c|}{parameter}
		&\multicolumn{1}{|c|}{best-fit}
		&\multicolumn{1}{|c|}{mean}
		&\multicolumn{1}{|c|}{std}\\ \hline
		$S(t_0)$ &   24.806 &   24.798 &   1.356  \\ \hline
		$E(t_0)$  &   16.123  &   15.292  &   0.781  \\ \hline
		$I(t_0)$  &  $1.35\times 10^{-3}$  &   0.092  &   0.076  \\ \hline
		$Z(t_0)$ &   0.333  &   0.517  &   0.452  \\ \hline
		$\epsilon$  &   0.995  &   0.976  &   0.077  \\ \hline
		$\beta$  &   2.365  &   2.341  &   0.115  \\ \hline
		$b$  &   0.077  &   0.378 &   0.351  \\ \hline
		$l$  &   0.365  &   0.406  &   0.227 \\ \hline
		$\mu$ &   0.031  &   0.036  &   0.009  \\ \hline
		$p$   &   0.007  &   0.068  &   0.051  \\ \hline
		$\rho$  &   3.897  &   4.008  &   0.461  \\ \hline
		$\Lambda$  &   11.553  &   12.033  &   0.634  \\ \hline
		$R_0^I$ &   74.821  &   65.245  &   13.808  \\ \hline
		
	\end{tabular}

	\begin{tabular}{||c|c|c|c||}\hline
		\multicolumn{4}{|c|}{\bf USSR} \\ \hline\hline
		\multicolumn{1}{|c|}{parameter}
		&\multicolumn{1}{|c|}{best-fit}
		&\multicolumn{1}{|c|}{mean}
		&\multicolumn{1}{|c|}{std}\\ \hline		
		$S(t_0)$  &   1.064  &   0.957 &   0.609  \\ \hline
		$E(t_0)$ &   4.129  &   2.660  &   0.481  \\ \hline
		$I(t_0)$ &   0.954  &   0.980  &   0.151  \\ \hline
		$Z(t_0)$   &   1.176  &   1.162  &   0.522  \\ \hline
		$\epsilon$  &   0.230  &   0.482  &   0.145  \\ \hline
		$\beta$ &   1.818  &   1.731  &   0.102  \\ \hline
		$b$  &   0.0112  &   0.267  &   0.187  \\ \hline
		$l$  &   0.730  &   0.649  &   0.247  \\ \hline
		$\mu$ &   0.075  &   0.070 &   0.023  \\ \hline
		$p$ &   0.097  &   0.104  &   0.071  \\ \hline
		$\rho$ &   3.340 &   3.341  &   0.506  \\ \hline
		$\Lambda$ &   18.134  &   18.288  &   1.785  \\ \hline
		$R_0^I$ &   18.806 &   25.055 &   10.614  \\ \hline
	\end{tabular}	
	\end{center}
	\caption{Parameter estimations for the SEIZ model and data for the
spread of Feynman diagrams for the USA, Japan, and the USSR. We restricted
the estimation procedure to the regime where $R_0^I>R_0^Z$, see
Eq.~(\ref{R0IZ}).}
\label{tabSEIZ}	
\end{table}

It is clear both from Table~\ref{tabdev} and from Fig.~\ref{figSEIZ} that
the SEIZ model gives the best fits to data, particularly in the case of
the USA.

We observed that good solutions (with very similar smallest deviation per
point) are possible with either idea strand $I$ or $Z$ having the largest
$R_0$. In this sense our initial six years of data cannot determine which
idea strand, adopters or skeptics, will eventually win out over many
generation times. In the parameter estimates presented in
Table~\ref{tabSEIZ} we have restricted the solutions to have
$R_0^I>R_0^Z$, thus limiting the search space to the historically
sanctioned eventual domination of Feynman diagrams over other techniques.  
This does not preclude the skeptics from growing initially in a population
of susceptibles, and we find in fact that degenerate solutions with and
without a growing number of skeptics are possible.

A novelty of the SEIZ model relative to the SEI is that the progression to
adoption can result from multiple contacts, both while susceptible
(parameterized by $\beta$) and while incubating (parameterized by $\rho$).
For every country the fact that $p$ is small and $\rho$ sizable makes
adoption favored and faster through contact with adopters while
incubating, relative to simple progression as in the SEI model. This may
indeed be the case in reality since the learning of Feynman diagrams in
the early years was characterized by extensive interpersonal contacts at
several stages of physicists' apprenticeship.  We know of only one case in
all three countries in which a few physicists learned about the diagrams
sufficiently well from articles or textbooks alone.  Practically every
adopter in all three countries is known to have interacted repeatedly with
other adopters before using the diagrams in their research \cite{DKDTA}.

\begin{figure}
\centerline{
\includegraphics[width=3.5in,angle=270]{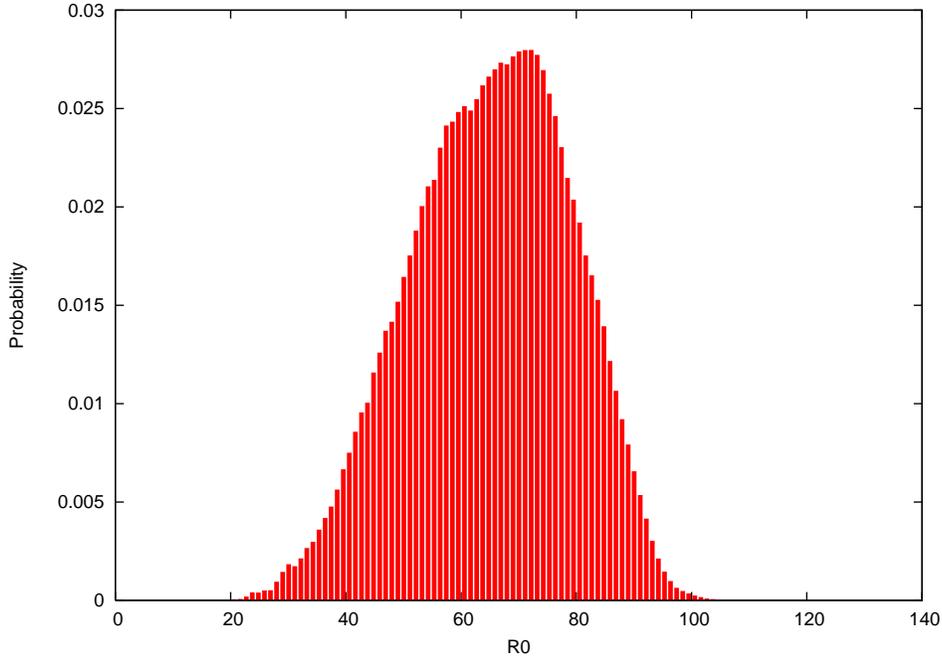}
}
\caption{The probability distribution of the basic reproductive number 
$R_0^I$ estimated from the SEIZ model for the Japan data on the spread of 
Feynman diagrams. The Japanese effectiveness of adoption is consistently 
greater than that for the other two countries, regardless of the specific 
model considered.}
\label{figR0_SEIZ}
\end{figure}

We also observe that for the SEIZ model the relative magnitudes of the
recruitment rates for the USA, USSR, and Japan follow the trends observed
in simpler models, while the same is approximately true also for the
effectiveness of adoption, $R_0$. The estimated probability distribution
function for $R_0^I$ for Japan in the SEIZ model is shown in
Fig.~\ref{figR0_SEIZ}.  As with the previous model, the large values of
$R_0^I$ estimated in the SEIZ model are mainly due to the very long
lifetime of the idea.

Among the models discussed above we are therefore inclined to prefer the
SEIZ model.  Not only does it best fit the empirical data, but it also
includes effects that we know to have been important, such as latency
(apprenticeship), adoption through multiple contacts, and institutional
and intellectual resistance.  Estimated parameters, both in their orders
of magnitude and (more important) in their relative sizes, reflect
properties of the idea's spread in each national community that match
qualitative expectations based on our empirical knowledge of the process.

\section{Conclusions}
\label{conclusions}
  
In this paper we applied several population models, inspired by
epidemiology, to the spread of a scientific idea, Feynman diagrams,  in
three different communities undergoing very different social transformations
during the middle years of the twentieth century.   
There is always a tradeoff between the use of models that include more detail (heterogeneous populations)  and highly aggregate simple models with a manageable number of parameters.  Although, a model built under very simplistic assumptions is expected to have deep limitations, the use of simple epidemic-type models have had tremendous success in the recent past, partly due to their ability to use existing data to make predictions (treatment for HIV \cite{penemaleho}) or recommendations  (control measures for SARS \cite{chfecaca}). This is the thinking behind the model choices made above. Moreover given the relative sparsity of quantitative data on social dynamical processes at present  such models may well prove to be the most useful starting points for modeling.
 
We have found that suitably adapted epidemic models do a good job of fitting the empirical data,
provided we allow their parameters to be very different from those
normally estimated for standard epidemics. In this sense the spread of
Feynman diagrams appears analogous to a very slowly spreading disease,
with characteristic progression times of years instead of days or weeks.
The spread of the diagrams also shows an enormous effectiveness of
adoption due primarily to the very long lifetime of the idea, rather than
to abnormally high contact rates.

The models give a quantification of parameters that are characteristic
both of the idea and of the mixing population in which it spreads. This
allows a more precise discussion of the sociological reasons why the idea
evolved differently in distinct national communities. The initial velocity
of spread of Feynman diagrams was fastest in Japan, followed by the USA,
and slowest in the USSR, probably as a result of geopolitical constraints
that severely limited access to the idea and its practitioners.  The
effectiveness of the adoption, encapsulated by $R_0$, was consistently
largest for Japan, most likely reflecting the high level of organization
of its scientific community in the difficult times that followed the end
of World War II. To our knowledge this is the first time that basic
reproductive number distributions have been estimated for the spread of an
idea. The USA and the USSR also show high recruitment rates, following the
two countries' massive investment in nuclear and high-energy physics
during the early Cold War. In this study we have done what seems to be yet uncommon in epidemiology, namely the estimation not only of model parameters and their variability,
 but also of the effective population sizes of the communities involved.

In the process of constructing epidemiological-type models and estimating
their parameters for the spread of Feynman diagrams, we had to confront
several conceptual issues concerning why the spread of ideas is or is not
analogous to that of a disease. One interesting aspect of the spread of
ideas is the inadequacy (or irrelevance) of the recovered state. In fact
many ideas may never be forgotten at all, as that would be in the worst
interest of the adopter.  As a result our parameter estimates consistently
find very long recovery times, $1/\gamma$. The same holds for the exit
rates, $1/\mu$.

In spite of these slow rates of exit and recovery, individuals commonly
have to acquire many ideas, and these may in some cases be mutually
exclusive, or at least may adversely affect the adoption of others.  We
introduced a new class of simple models with multiple $Z$ classes
representing these strands. It is a curious, and we believe important fact
that the recruitment of individuals from a class of susceptibles to other
ideas has the same mathematical effect as vaccination against disease. In
this sense ``immunity'' to an idea may be obtained either by education
about its possible implications (perhaps analogous to actual
immunization), or by distraction with other, more easily acquired concepts
embodied by the $Z$ classes.

We must emphasize that the behavior of individuals when exposed to
ideas may be very different, indeed opposite, to what they may do during
an epidemic outbreak. First, people intentionally seek ways to extend the
infectious period of an idea, usually by recording it and storing it in
various documents. In this sense the lifetime of an idea can largely
transcend that of individuals. Second, short of vaccination the most
effective strategy to stop a disease epidemic is through isolation, which
reduces the contact rate.  Ideas, unlike diseases, are usually beneficial
and thus people's behavior tends to maximize effective contacts. This
pattern can be captured through the mapping of the social network of
contacts that underlie the spread of the idea, which we analyze elsewhere
\cite{BettencourtKaiser}. There we show that the communities where Feynman
diagrams spread the fastest had created intentional social and behavioral
structures that ensured very efficient communication of scientific
knowledge.

We finish by remarking that the SEIZ model, which included both skeptic and incubator classes, as well as acceleration to adoption from incubation (parameterized by $\rho$), captures most adequately the role of such classes in the transmission process, since it yields the best fits (smallest average deviations in Table ~\ref{tabdev}).  Nevertheless the modeling of the spread of ideas  discussed above is but a  simple caricature of the complex social dynamical processes involved.  Our hope is that this work may bring a new and hopefully useful quantitative perspective into the study of the diffusion of ideas, by the simplest means possible.

\section*{Acknowledgements}

	We thank Gerardo Chowell, Ed MacKerrow, Miriam Nu\~{n}o and Steve
Tennenbaum, for discussions and comments.   A. Cintr\'{o}n-Arias  
acknowledges financial support from Mathematical and Theoretical Biology Institute 
and Center for Nonlinear Studies at Los Alamos National Laboratory.  Collaboration was greatly facilitated through visits by several of the authors to the  Statistical and Applied Mathematical Sciences Institute (SAMSI), Research Triangle Park,  NC, which is funded by NSF under grant DMS-011209.  The authors thank the hospitality of Santa Fe Institute, where portions of this work were undertaken.

                \appendix
 \section*{Appendix A. Ensemble parameter estimation procedure}
 
Here we give a short description of our parameter estimation procedure
and uncertainty quantification.

The problem of generating estimates for model parameters describing the
spread of ideas is the absence of clear quantitative expectations, both
concerning which model should apply best and what the quantitative value
of its parameters should be. As such we devised a novel search method
capable of both finding the best fit to the data possible given a choice
of model, but also of producing an ensemble of solutions that are
compatible with the data within a certain admissible error.

As a starting point we take the fact that simple population models cannot
be expected to give perfect descriptions of the data, resulting in a
minimum level of discrepancy. We chose to parameterize this discrepancy by
a collective measure of the average absolute value of the deviation
between the best model prediction and each data point. This measure allows
us to discuss and compare how good models are at describing a specific
data set. Our results are given in Table \ref{tabdev}.

Second, we expect in general that data contains errors, e.g. early
underestimation, false positives, accounting errors. Thus a given level of
uncertainty in the data will translate into parameter statistical
distributions that are compatible with those allowable deviations.  This
is a stochastic optimization problem (see, e.g., \cite{stochopt} for a
general discussion).  Based on this idea we perform an estimation of the
joint parameter distribution of model parameters, conditional on a set of
allowable deviations at each datum. To be specific we can write that the
unknown exact data point $I^E(t_i)$, measured at time $t=t_i$, can be
written in terms of the observed datum $I^O(t_i)$ and an error $\xi(t_i)$
as
\begin{equation}
 I^E(t_i) = I^O(t_i)+ \xi(t_i).
 \end{equation}
The error $\xi(t_i)$ is only known statistically. In order to proceed we
must specify a model for $\xi$. Here we assumed a simple Gaussian
distribution such that
\begin{equation}
 P\left[ \xi(t_i) \right]=P \left[ I^E(t_i) - I^O(t_i) \right] = {\cal N} 
e^{-\frac{\xi^2(t_i)}{2 \sigma^2(t_i)} },
\end{equation}
where $\cal N$ is the normalization factor and $\sigma(t_i)$ 
parameterizes the expected error at time $t=t_i$.

This expectation for the errors can be translated into a commensurate
fitness function (analogous to a Hamiltonian in statistical physics) that
can in turn be minimized in order to produce parameter estimates through a
search procedure.  For each model realization (in terms of a set of
parameters ${\cal S}=(S(t=t_0),E(t=t_0), ...,\beta, \gamma,...)$) we take
this function to be
\begin{eqnarray}
H({\cal S})= \sum_i \frac{ \left[ I^M(t_i) - I^O(t_i) \right] ^2}{2 
\sigma^2(t_i)} ,
\end{eqnarray}
which is an implicit function of $\cal S$. 
If the model could generate exact results we could then make the natural
association $I^E(t_i) \rightarrow I^M(t_i)$. This is usually not the case, 
since
a residual minimal deviation always persists.  To account for this we
normalize this function to zero by taking $H'({\cal S})=H-H_0$, i.e. by
subtracting the minimal value of $H$, obtained for the best parameter set.

Given this choice of $H'$ we can produce, in analogy with standard
procedures in statistical physics, a joint probability distribution for
model parameters given by
\begin{eqnarray}
P({\cal S}) \sim e^{-H'} .
\label{prob}
\end{eqnarray}
This choice guarantees that all statistical moments are finite.  This
joint probability distribution can then be used to compute any moment of
any set of parameters, including single parameter distribution functions,
and cross-parameter correlations such as covariances. In Section
\ref{results} we show results for the single parameter averages and their
standard deviations. We also show some single parameter probability
distribution functions.

In general the estimation of this probability distribution can be obtained
by randomly generating many model parameter sets and weighing them
according to Eq.~(\ref{prob}). The procedure is slightly complicated
because we are dealing with an inverse problem in which, given a trial set
of parameters, comparison with the data is performed only after the
non-linear model dynamical equations have been solved. Fortunately for
models that consist of small numbers of ordinary differential equations
the computational effort is not prohibitive.

In practice we used an ensemble of trial solutions, from which we select a
number of best strings, according to a standard Monte Carlo procedure,
weighted by Eq.~(\ref{prob}), to generate the next generation of the
ensemble. In order to do this we introduce a mutation implemented in terms
of random Gaussian noise around the best parameter sets. This yields an
effective minimization method, capable of exploring large regions of
parameter space. It also creates as a byproduct an ensemble of good
strings with small deviations to the data. For small enough deviations
from the best string we can sample parameter space in an unbiased manner.
It is this ensemble, and its best string, that is then used to estimate
Eq.~(\ref{prob}). Results given in Section \ref{results} involve ensembles
with several million realizations and a choice of $\sigma$, common to all
points, corresponding to $10\%$ deviation between the best parameter
estimate and other ensemble members.


\begin{thebibliography}{}
		
	 \bibitem{may} R. M. May, in
	\newblock{\em Theoretical Ecology: Principles and Applications, 2nd edn,} 
	\newblock{edited by R.M. May (Sinauer, Sunderland, 1981).}

	
	\bibitem{yod}P. Yodzis,
	\newblock{\em Introduction to Theoretical Ecology}
	\newblock{(Harper \& Row Publishers, New York, 1989).} 
	
						
	\bibitem{thi} H. Thieme, 
	\newblock{\em Mathematics in Population Biology}
	\newblock{(Princeton University Press, Princeton, 2003). }
	
	\bibitem{anma}R. Anderson and R. May, 
	\newblock{\em Infectious Diseases of Humans: Dynamics and Control}
	\newblock{(Oxford University Press, Oxford, 1991)}.

	\bibitem{dihe}O. Diekmann and J. A. P. Heesterbeek,
	\newblock{{\em Mathematical Epidemiology of Infectious Diseases: 
Model Building, Analysis and Interpretation}}
	\newblock{(Wiley, New York, 2000)}.


	  \bibitem{alli} L. Allen,
           \newblock{{\em An Introduction to Stochastic Processes with 
Applications to Biology}}
           \newblock{(Pearson Education-Prentice Hall, New Jersey, 2003)}.

	
	\bibitem{brca} F. Brauer and C. Castillo-Ch\'{a}vez, 
	\newblock{{\em Mathematical Models in Population Biology and 
	Epidemiology}}
	\newblock{(Springer-Verlag, New York, 2001).} 
       
	
	\bibitem{penemaleho} A. S. Perelson {\em et al.,}
	\newblock{{\em Science}, {\bf 271}, 5255, 1582-1586 (1996).}		
	
			
	\bibitem{hucoca} W. Huang, K. Cooke and C. Castillo-Ch\'{a}vez,
	\newblock{{\em SIAM J.  Appl. Math.}, {\bf 52},  835-854 (1990)}.
         
         
	\bibitem{watt} D. Watts,
	\newblock{{\em P.  Natl. Acad. Sci. USA}, {\bf 99}, 5766-5771 
(2002)}.	


	\bibitem{bihiwe}S. Bikhchandani, D. Hirshleifer and I. Welch,  
	\newblock{{\em J.  Polit. Econ.}, {\bf 100},  992-1026 (1992)}.


	\bibitem{bet} L. M. A. Bettencourt,     
	\newblock{\texttt http://xxx.lanl.gov/abs/cond-mat/?0212267}
	
	

	\bibitem{stma}D. Strang and M. Macy,
	\newblock{{\em Am. J. Sociol.}, {\bf 107},  147-182 (2001)}.
	
	
	  \bibitem{rog}E. Rogers,
            \newblock{\em Diffusion of Innovations}
             \newblock{(Free Press, New York, 1995)}.
	

	\bibitem{caso} C. Castillo-Ch\'{a}vez and B. Song, in
	\newblock{{\em Bioterrorism: Mathematical Modeling Applications in 
Homeland Security, SIAM Frontiers in Applied Mathematics}, edited by H.T. 
Banks and C. Castillo-Ch\'{a}vez
	(SIAM, Philadelphia, 2003), Vol. 28, p.155}.

	\bibitem{sancas} F. S\'{a}nchez-Pe\~{n}a and C. 
Castillo-Ch\'{a}vez,
	\newblock{in preparation}.

	 \bibitem{ghovk}B. Gonz\'{a}lez {\em  et al.},
                \newblock{{\em J. Math. Psychol.}, {\bf 47}, 515-526 
(2003).}               


	  \bibitem{rap}A. Rapoport,
                  \newblock{{\em Bull. Math. Biophys.}, {\bf 15}, 
523--533 (1953).}
                                                       
	  \bibitem{dake1}D. J. Daley and D. G. Kendall,
                     \newblock{{\em J. I. Math. Appl.},  {\bf 1}, 42--55 
(1965). }
                           
              \bibitem{gof} W. Goffman,
              \newblock{{\em Nature}, {\bf 212}, 5061, 449-452 (1966).}
              
                                 
             \bibitem{tab} 
             \newblock{For a recent review see A. N. Tabah, 
              in {\em Annual Review of Information Science and Technology (ASIS)},
		edited by M. E. Williams (Information Today, Medford, 1999), Vol. 34, p. 249}.
             
   
                                 
          \bibitem{adhu} L. Adamic and B. Huberman, in
            \newblock{{\em Complex Networks, Lecture Notes in Physics},
		edited by E. Ben-Naim, H. Frauenfelder and Z. Toroczkai
		 (Springer, Berlin, 2004), Vol. 650, p. 371}.

	 \bibitem{adad}E. Adar, Z. Li, L. A. Adamic and R. Lukose, in
         \newblock{{\em Workshop on the Weblogging Ecosystem, 13th
       International World Wide Web Conference, New York, 2004}; L.A. 
Adamic and E. Adar, {\em Soc. Networks}, {\bf 25}, 3, 211-230 (2003)}.

	 \bibitem{kekle} D. Kempe and J. Kleinberg, in
	\newblock{{\em Proceedings 43rd Symposium on Foundations of 
Computer Science},
        (IEEE Computer Society, Los Alamitos, 
2002), p. 471-480.}
                        
	\bibitem{kemc} W. Kermack and A. McKendrick, 
	\newblock{{\em P. R. Soc. Lon. Ser. A}, {\bf 115}, 772,700-721 
(1927)}.
    
                                
              \bibitem{het} H. Hethcote,
              \newblock{{\em SIAM Rev.}, {\bf 42},  599-653  (2000)}.
       
       \bibitem{mopave} Y. Moreno, R. Pastor-Satorras and A. Vespignani, 
	\newblock{{\em Eur. Phys. J. B}, {\bf 26}, 521-529 (2002).}

            \bibitem{ros0} R. L. Rosnow,
            \newblock{{\em Am. Psychol.}, {\bf 46},   484-495  (1991)}.
         
          \bibitem{bod1} P. Bordia and N. DiFonzo,
         \newblock{{\em Asian J. Soc. Psychol.}, {\bf 5}, 49-61 (2002).}
         
        \bibitem{brre} J. J. Brown and P. H. Reingen,
        \newblock{{\em J. Consum. Res.}, {\bf 14}, 350-362 (1987).}
            
        \bibitem{keklta} D. Kempe, J. Kleinberg and E. Tardos, in
        \newblock{{\em Proc. 9th ACM SIGKDD Intl. Conf. on Knowledge
Discovery and Data Mining}, (ACM, New York, 2003).}

	  \bibitem{dipe} R. E. Dickinson and C. E. M. Pearce,
          \newblock{ {\em Math. Comput. Model.}, {\bf 38}, 1157-1167 
(2003).}

	    \bibitem{zan1} D. H. Zanette,
            \newblock{{\em Phys. Rev. E}, {\bf 65}, 041908 (2002).}          
	
	 \bibitem{mnp} Y. Moreno, M. Nekovee and A. F. Pacheco,
        \newblock{{\em Phys. Rev. E},  {\bf 69}, 066130  (2004)}.
                                 
	 \bibitem{tcd} K. Thompson {\em et al.},
         \newblock{Mathematical and Theoretical Biology Institute 
Technical Report, 2003 (unpublished)}.

	\bibitem{DKDTA} D. I. Kaiser,
	\newblock{\it Drawing Theories Apart:  The Dispersion of Feynman 
Diagrams in Postwar Physics}
          \newblock{(University of Chicago Press, Chicago, 2005).}
			
 	\bibitem{KIH} D. I. Kaiser, K. Ito, and K. Hall,
	\newblock{Soc. Stud. Sci., {\bf 34}, 6, 879-922 (2004).}

	
	\bibitem{BettencourtKaiser} L. M. A.  Bettencourt and D. I. Kaiser, in preparation.
	
	\bibitem{DKHSPS} D. I. Kaiser,
	\newblock{ Hist. Stud. Phys. Bio. Scis. {\bf 33}, 1, 131-159 (2002).}


	 \bibitem{Schweber} S. S. Schweber, 
	 \newblock{\it QED and the Men Who Made It: Dyson, Feynman, 
Schwinger, and Tomonaga} 
	\newblock{(Princeton University Press, Princeton, 1994).}


 	\bibitem{incub} It is customary to model the movements out of the
class $E$ into the next class $I$ by a term like $\epsilon E$.  This
corresponds to having exponentially distributed waiting times in the $E$
class.  In other words, the simple progression rate $\epsilon E$
corresponds to $\mathcal{P}(\tau)=\exp(-\epsilon \tau)$ as the fraction
that is still in the incubator class $\tau$ units after entering this
class, and to $1/\epsilon$ as the mean waiting time.
	
	\bibitem{casfenhua1} C. Castillo-Ch\'{a}vez, Z. Feng and W. 
Huang, in
	 \newblock{{\em Mathematical Approaches for Emerging and 
Reemerging Infectious Diseases, 
	 The IMA Volumes in Mathematics and its Applications}}, 
	 \newblock{edited by  C. Castillo-Ch\'{a}vez {\em et al.}
	 (Springer, New York, 2002), Vol. 125, p. 229.}

	 \bibitem{vadwa1} P. van den Driessche and J. Watmough,
	\newblock{{\em Math. Biosci.}, {\bf 180},  29-48 (2002).}	
			
	\bibitem{chfecaca} G. Chowell, P. W. Fenimore, M. A. 
Castillo-Garsow and C. Castillo-Ch\'{a}vez,
	\newblock{{\em J. Theor. Biol.}, {\bf 224}, 1-8 (2003).}
	

	  \bibitem{DysonNature} F. Dyson,  
                  \newblock{ {\em Nature}, {\bf 427}, 297 (2004)}.

	\bibitem{stochopt} J. C. Spall,
	\newblock{\em Introduction to stochastic search and optimization}.
	\newblock{(Wiley- Interscience, Hoboken, New Jersey, 2003). }  


        \end{thebibliography}
\end{document}